\documentclass[fleqn,usenatbib]{mnras}

\usepackage{newtxtext} 
\usepackage{mathptmx}
\usepackage{txfonts}

\usepackage[T1]{fontenc}
\usepackage{xcolor} 
\DeclareRobustCommand{\VAN}[3]{#2}
\let\VANthebibliography\thebibliography
\def\thebibliography{\DeclareRobustCommand{\VAN}[3]{##3}\VANthebibliography}

\usepackage{graphicx}	
\usepackage{amssymb}	

\DeclareRobustCommand{\ion}[2]{%
\relax\ifmmode
\ifx\testbx\f@series
{\mathbf{#1\,\mathsc{#2}}}\else
{\mathrm{#1\,\mathsc{#2}}}\fi
\else\textup{#1\,{\mdseries\textsc{#2}}}%
\fi}

\title[Nature of Infrared Bright, Optically Dark Galaxies with JWST]{Unveiling the Nature of Infrared Bright, Optically Dark Galaxies with Early JWST Data}

\author[L. Barrufet et al.]{L. Barrufet$^{1}$\thanks{E-mail:laia.barrufetdesoto@unige.ch}, 
P. A. Oesch$^{1,2}$, 
A. Weibel$^{1}$, 
G. Brammer$^{2}$,
R. Bezanson$^{3}$,
R. Bouwens$^{4}$,
Y. Fudamoto$^{5,6}$, 
\and
V. Gonzalez$^{7}$,
R. Gottumukkala$^{1}$,
G. Illingworth$^{8}$, 
K. E. Heintz$^{2}$, 
B. Holden$^{8}$, 
I. Labbe$^{9}$,
D. Magee$^{8}$,
\and
R. P. Naidu$^{10}$,
E. Nelson$^{11}$, 
M. Stefanon$^{12, 13}$,
R. Smit$^{14}$,
P. van Dokkum$^{15}$,
J. R. Weaver$^{16}$,
C. C. Williams$^{17, 18}$
\\ 
$^{1}$Department of Astronomy, University of Geneva, Chemin Pegasi 51, 1290 Versoix, Switzerland \\
$^{2}$Cosmic Dawn Center (DAWN), Niels Bohr Institute, University of Copenhagen, Jagtvej 128, K\o benhavn N, DK-2200, Denmark\\
$^{3}$Department of Physics and Astronomy and PITT PACC, University of Pittsburgh, Pittsburgh, PA 15260, USA\\
$^{4}$Leiden Observatory, Leiden University, PO Box 9500, 2300 RA Leiden, The Netherlands  \\
$^{5}$Waseda Research Institute for Science and Engineering, Faculty of Science and Engineering, Waseda University, 3-4-1 Okubo, Shinjuku, Tokyo 169-8555, Japan \\
$^{6}$National Astronomical Observatory of Japan, 2-21-1, Osawa, Mitaka, Tokyo, Japan \\
$^{7}$Departamento de Astronomia, Universidad de Chile, Camino del Observatorio 1515, Las Condes, Santiago 7591245, Chile \\
$^{8}$Department of Astronomy and Astrophysics, University of California, Santa Cruz, CA 95064, USA \\
$^{9}$Centre for Astrophysics and Supercomputing, Swinburne University of Technology, Melbourne, VIC 3122, Australia \\
$^{10}$Center for Astrophysics $|$ Harvard \& Smithsonian, 60 Garden Street, Cambridge, MA 02138, USA \\
$^{11}$Department for Astrophysical and Planetary Science, University of Colorado, Boulder, CO 80309, USA \\
$^{12}$Departament d'Astronomia i Astrofisica, Universitat de Valencia, C. Dr. Moliner 50, E-46100 Burjassot, Valencia,  Spain \\
$^{13}$Unidad Asociada CSIC "Grupo de Astrofisica Extragalactica y Cosmologia" (Instituto de Fisica de Cantabria - Universitat de Valencia) \\
$^{14}$Astrophysics Research Institute, Liverpool John Moores University, 146 Brownlow Hill, Liverpool L3 5RF, UK \\
$^{15}$Astronomy Department, Yale University, 52 Hillhouse Avenue, New Haven, CT 06511, USA \\
$^{16}$Department of Astronomy, University of Massachusetts, Amherst, MA 01003, USA \\
$^{17}$ NSF's National Optical-Infrared Astronomy Research Laboratory, 950 North Cherry Avenue, Tucson, AZ 85719, USA\\
$^{18}$Steward Observatory, University of Arizona, 933 N. Cherry Avenue, Tucson, 85721, USA  \\
}

\date{Accepted XXX. Received YYY; in original form ZZZ}

\pubyear{2022}

\begin{document}
\label{firstpage}
\pagerange{\pageref{firstpage}--\pageref{lastpage}}
\maketitle

\begin{abstract} 
Over the last few years, both ALMA and Spitzer/IRAC observations have revealed a population of likely massive galaxies at $z>3$ that was too faint to be detected in HST rest-frame ultraviolet imaging. However, due to the very limited photometry for individual galaxies, the true nature of these so-called HST-dark galaxies has remained elusive. Here, we present the first sample of such galaxies observed with very deep, high-resolution NIRCam imaging from the Early Release Science Program CEERS. 30 HST-dark sources are selected based on their red colours across 1.6 $\mu$m to 4.4 $\mu$m. Their physical properties are derived from 12-band multi-wavelength photometry, including ancillary HST imaging. We find that these galaxies are generally heavily dust-obscured ($A_{V}\sim2$ mag), massive ($\log (M/M_{\odot}) \sim10$), star-forming sources at $z\sim2-8$ with an observed surface density of  $\sim0.8$ arcmin$^{-2}$. This suggests that an important fraction of massive galaxies may have been missing from our cosmic census at $z>3$ all the way into the Reionization epoch.  The HST-dark sources lie on the main sequence of galaxies and add an obscured star formation rate density (SFRD) of $\mathrm{3.2^{+1.8}_{-1.3} \times 10^{-3} M_{\odot}/yr/Mpc^{3}}$ at $z\sim7$ showing likely presence of dust in the Epoch of Reionization. 
Our analysis shows the unique power of JWST to reveal this previously missing galaxy population and to provide a more complete census of galaxies at $z=2-8$ based on rest-frame optical imaging.


\end{abstract}


\newcommand{\TBD}[1]{\textcolor{red}{\bf{#1}}}	

\section{Introduction}

By $z>3$, the rest-frame optical light redshifts out of the view of the Hubble Space Telescope (HST). Therefore, our understanding of `normal' star-forming galaxies such as Lyman Break Galaxies (LBGs) that were identified in HST strongly relies on rest-frame ultra-violet (UV) observations. This can result in an incomplete galaxy census at earlier times, due to UV-faint galaxy populations such as quiescent or dust-obscured sources.

While a small number of extremely obscured, star-forming galaxies (e.g., sub-millimetre galaxies; SMGs) have been known to exist at $z>4$ for a long time, such galaxies are 100$\times$ less common than LBGs (with a sky density of only 0.01 arcmin$^{-2}$; e.g., \citealt{Riechers2013}; \citealt{Marrone2018}). Hence, they only provide a limited contribution to the integrated star formation rate density (SFRD). Similarly, massive ($\mathrm{>10^{11}M_{\odot}}$) quiescent objects have only been identified to $z\sim3-4$ 
\citep[e.g.][]{Tanaka2019, Valentino2020, Carnall2020, Santini2021}, but they remain relatively rare such that degree-scale survey volumes are needed for their detection. Nevertheless, due to our limited sensitivity in rest-frame optical light at $z>3$, it is still possible that the early cosmic star formation and stellar mass densities are dominated by less extreme, dusty star-forming galaxies (DSFGs) or quiescent galaxies, which have been missing from rest-UV datasets. The key question is: \textit{how common are such red sources at $z > 3$?}

Over the last few years, evidence has been mounting that, indeed, a population of obscured, star-forming galaxies with less extreme SFRs than SMGs could be quite common in the early Universe. These sources have been undetected in current HST surveys, due to their very red colours between $1.6\,\mu$m and $4.5\,\mu$m (e.g., \citealt{Franco2018},  \citealt{Wang2019}), and they have thus been named HIEROs (HST to IRAC extremely red objects), H-dropouts,  HST-dark galaxies, or HST-faint. Very red dusty sources exist even below the sensitivity limits in IRAC, making them only detectable by ALMA. However, previous studies with ALMA relied on small areas of the sky, thus reporting only one or two galaxies in some cases \citep[e.g.,][]{Williams2019, Fudamoto2021}.
Hence, HST-dark sources are typically selected in Spitzer/IRAC imaging or as serendipitous sources in ALMA sub-mm continuum data \citep[e.g.,][]{Huang2011, Caputi2015, Simpson2014, Stefanon2015, Wang2016, Franco2018, AlcaldePampliega2019, Wang2019, Yamaguchi2019, Williams2019, Umehata2020, Sun2021, Fudamoto2021, Manning2022}. However, due to the limited photometric information available, their spectral energy distributions (SEDs) and redshifts remained uncertain. Nevertheless, the general consensus in the literature is that the vast majority of these galaxies are heavily dust-obscured, star-forming sources at $z\sim3-6$ that dominate the high-mass end of the galaxy population at these early cosmic times \citep[e.g.,][]{AlcaldePampliega2019,Wang2019}. 

Accurately constraining the number densities and the physical properties of such sources is critical for our understanding of early galaxy build-up. It is conceivable that dust-obscured, HST-undetected sources dominate the SFRD at $z>4$ unlike what is typically assumed \citep[e.g.,][]{Williams2019, Gruppioni2020, Zavala2021}. However, the most recent 2mm number counts as well as IR luminosity functions measured from ALMA continuum observations out to $z\sim4-7$ indicate a less extreme scenario \citep[see][Barrufet submitted]{Casey2018}. Nevertheless, it has become clear that our census of galaxies at $z>3$ is indeed far from complete, leaving a critical gap in our understanding of early galaxy build-up. 

Finding and characterizing this missing galaxy population is a key goal of extragalactic astronomy in order to inform theoretical models of galaxy evolution. In this paper, we address this important question with some of the first JWST NIRCam imaging from the Early Release Science (ERS) program CEERS \citep{Finkelstein2022}. This allows us to identify HST-dark galaxies with the same selection that has been used in the past for HST+Spitzer samples, but now with significantly measured multi-wavelength photometry in several bands to derive their redshifts and physical properties reliably for the first time. This follows previous analyses with JWST data that have already revealed massive galaxies at $z>7$ \citep[see][]{Santini22, Labbe22}. In a separate paper, we also analyze the morphologies of bright F444W selected sources \citep{Nelson2022}.

This paper is organized as follows.
In Section \ref{observations} we describe the observational data and sample selection.  
Section \ref{modelingandaverageprop} outlines our spectral energy distribution modelling and the variety of physical properties that we find for HST-dark galaxies. 
In Section \ref{results}, we present our results and compare our galaxy sample to SMGs, before we derive their 
contribution to the CSFRD in Section \ref{SFRDsection} and finish with a summary and conclusions in Section \ref{Summary}. 

Magnitudes are given in the AB system \citep[e.g.,][]{Oke83}, and where necessary we adopt a \citet{Planck2015} cosmology.

\section{Observations and Sample Selection}
\label{observations}

\subsection{CEERS observations and HST+JWST photometric catalogue}

This analysis is based on the first four NIRCam pointings from the Early Release Science (ERS) Program CEERS (PI: Finkelstein; PID: 1345; see \citealt{Finkelstein2022}). 
These images include exposure sets of short and long-wavelength filters: F115W+F277W, F115W+F356W, F150W+F410M, and F200W+F444W. The typical integration time for each filter set is 0.8 hr. Hence, the F115W images obtained double the exposure time compared to the remaining filters. 

\begin{figure}
 \centering
  \includegraphics[width=\columnwidth]{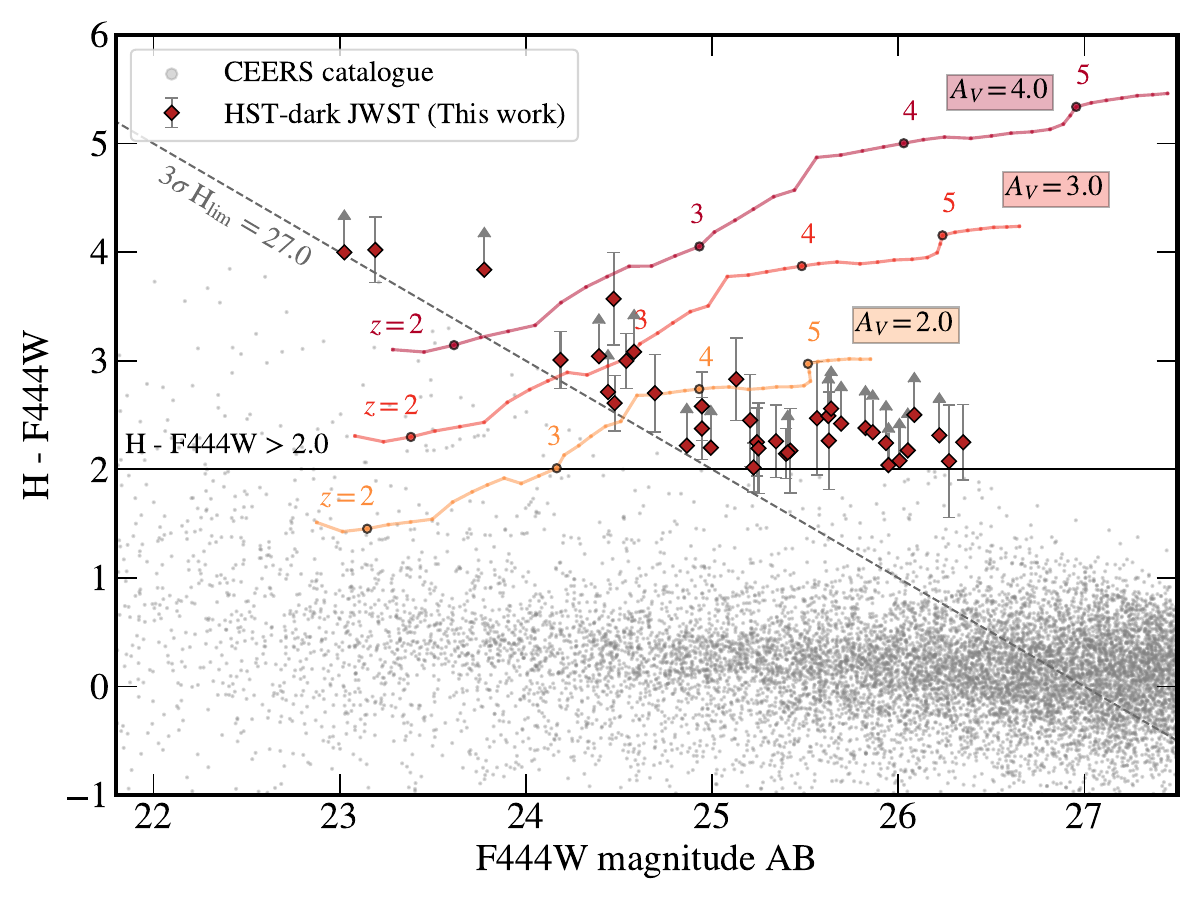}
 \caption{ Colour-magnitude diagram (F160W-F444W vs F444W) used to select the HST-dark galaxies. The dashed line shows the location of $\mathrm{H = 27}$ mag, which corresponds to the average $3\sigma$ limit of the shallowest parts of the H-band in the HST data. The black line marks the adopted colour cut $\mathrm{H-F444W > 2.0}$ and the red points correspond to the galaxies included in our sample. Of the 39 HST-dark galaxies selected, 19 have lower limit fluxes in the H-band (grey arrows). The continuous lines show the theoretical evolution in galaxy colours at redshifts of $\mathrm{2<z<6}$ for three different levels of reddening $\mathrm{A_{v} = 2.0, \ 3.0,  \ 4.0}$ mag (yellow, orange and purple lines  respectively).  The grey dots represent the density of sources from the total CEERS catalogue with detections in the F444W band ($\mathrm{\sim 26,000}$ sources).  } 
  \label{colour_selection}
\end{figure}

The data were retrieved from the MAST archive and were then WCS-matched and combined using the publicly available \texttt{grizli}\footnote{\url{https://github.com/gbrammer/grizli/}} pipeline. Most notably, the images are aligned with the Gaia DR3 catalogues. For details on the reduction, we refer the reader to Brammer et al., in prep. In the following, we work with images at 40mas pixel scale. 
The 5$\sigma$ depths are measured in circular apertures of 0\farcs48 diameter. They range from 27.8 mag in the medium band filter F410M to 28.4-28.6 mag in the wide filters F200W, F277W and F356W at $2-3.5\,\mu$m (see also Table 1 in \citealt{Naidu22}).

\begin{figure*}
 \centering
   \includegraphics[width=\columnwidth]{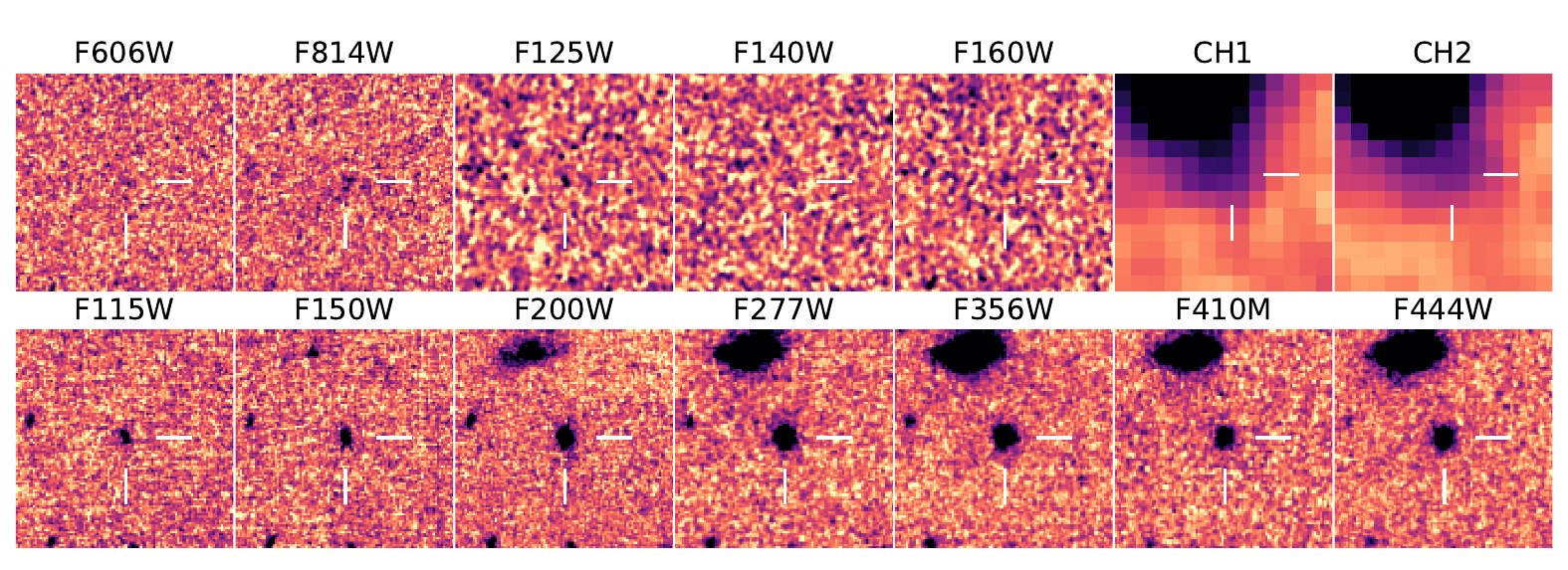} 
     \includegraphics[width=\columnwidth]{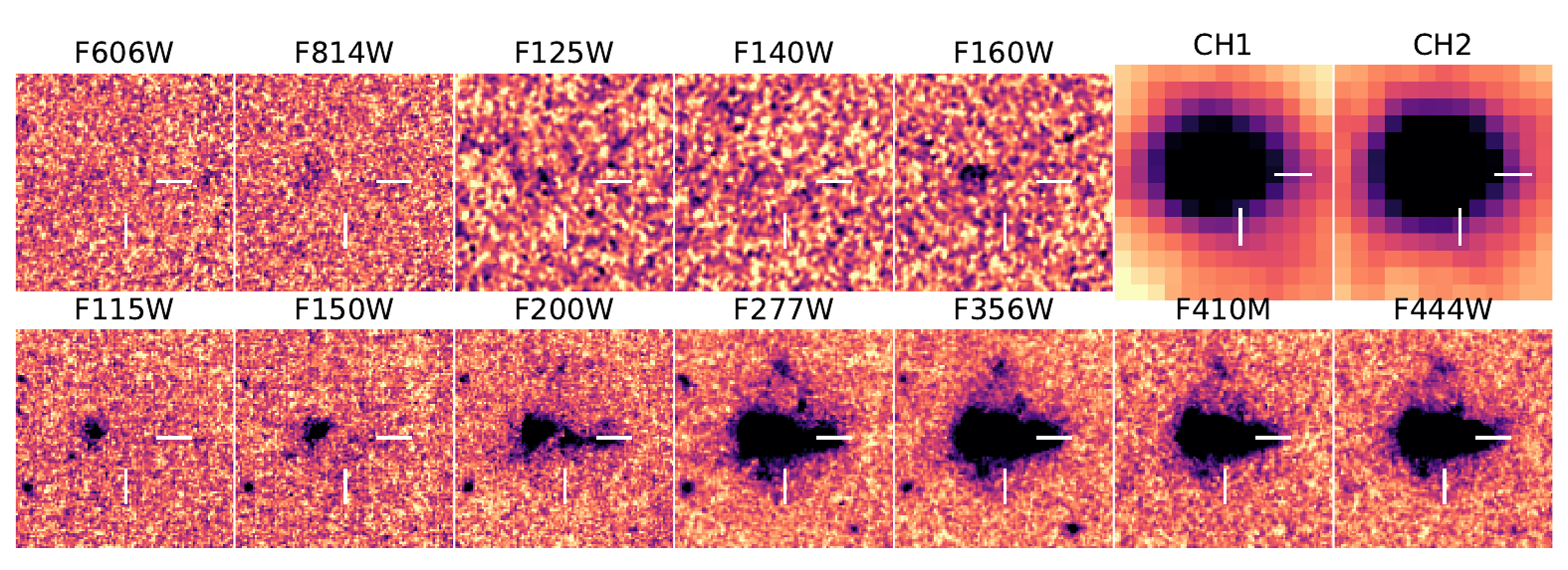} 

   \includegraphics[width=\columnwidth]{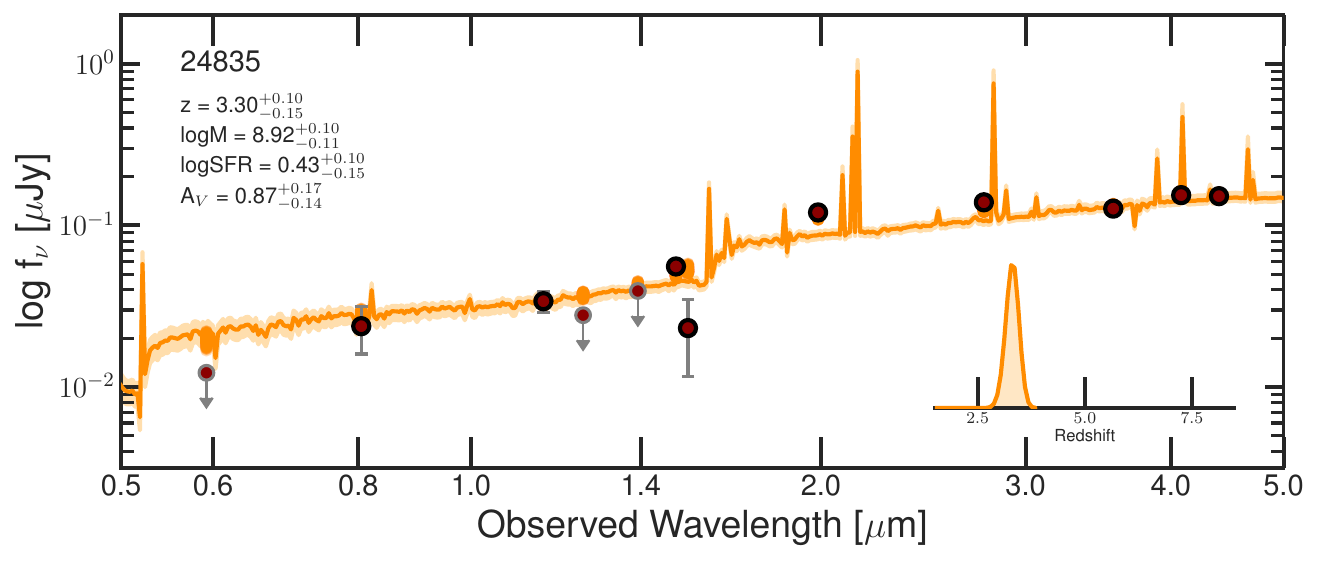} 
      \includegraphics[width=\columnwidth]{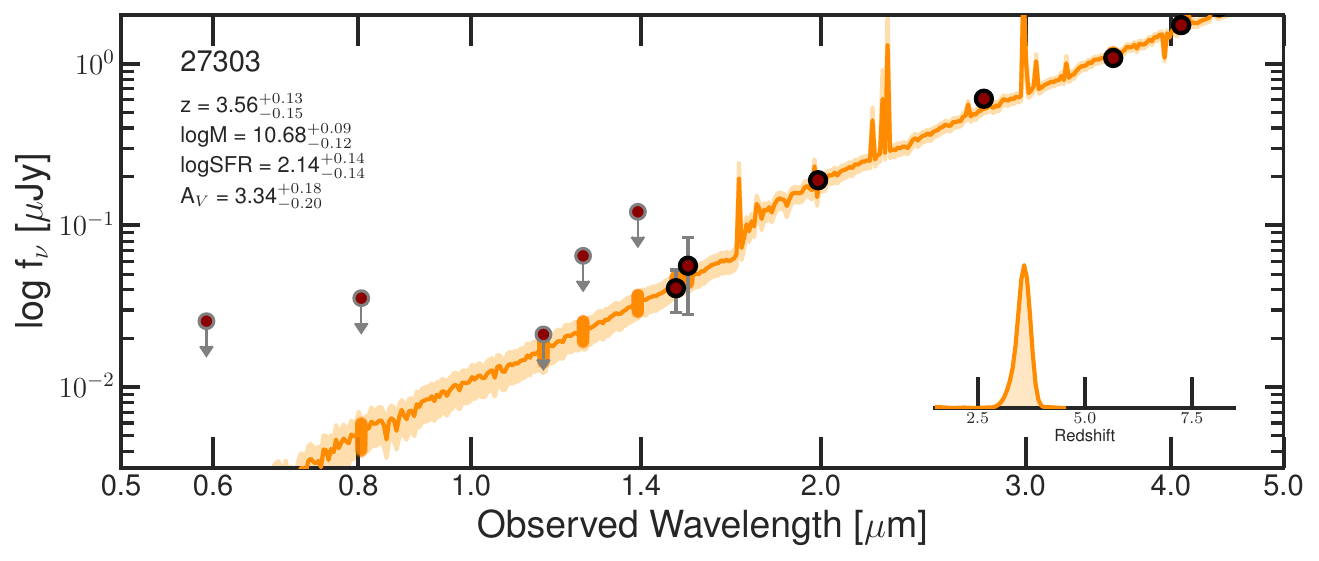}

    \includegraphics[width=\columnwidth]{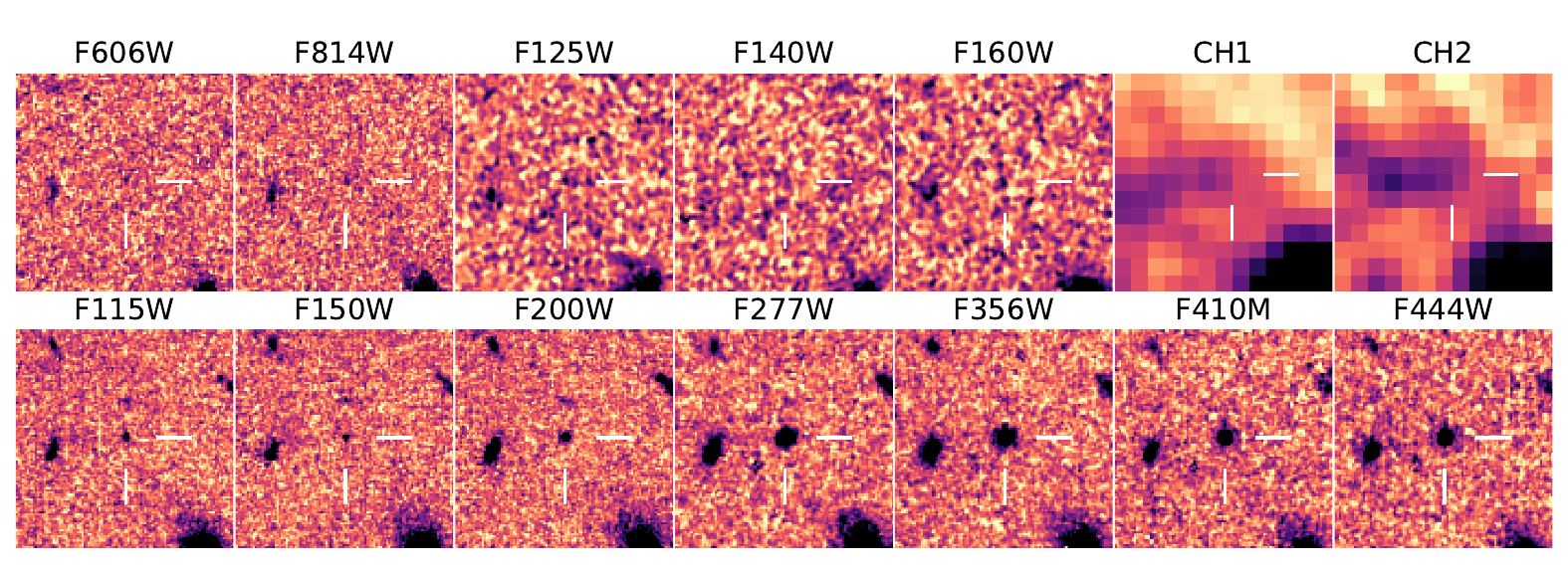} 
       \includegraphics[width=\columnwidth]{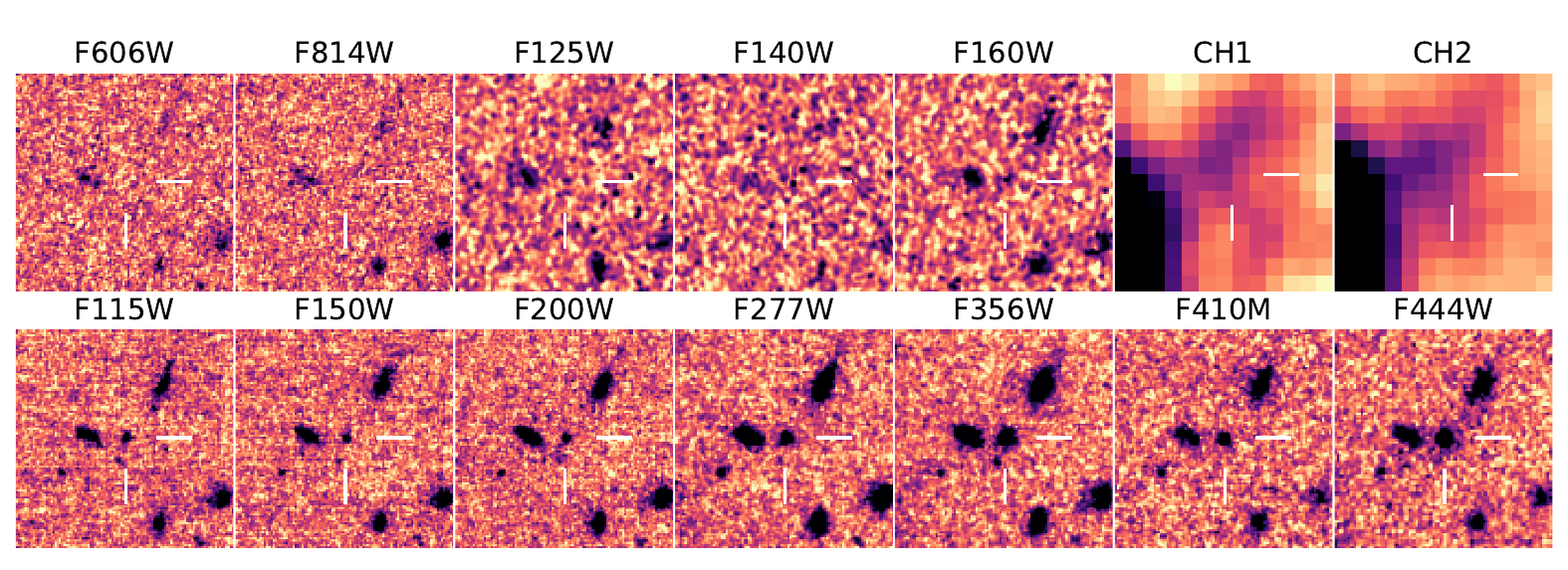} 

    \includegraphics[width=\columnwidth]{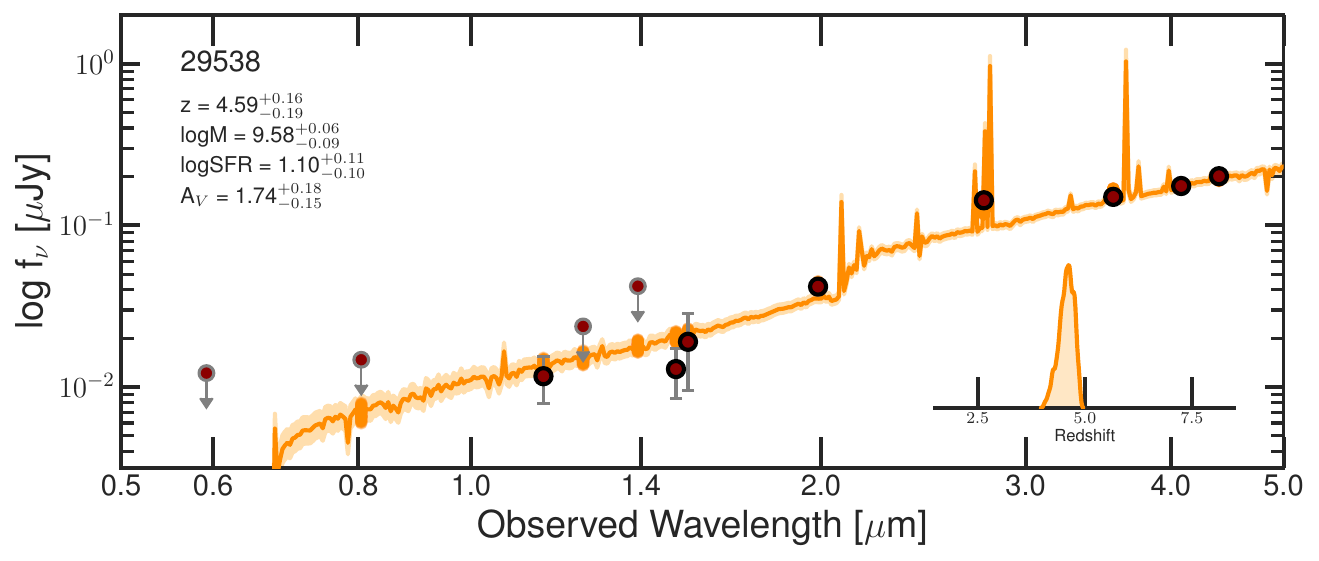}
       \includegraphics[width=\columnwidth]{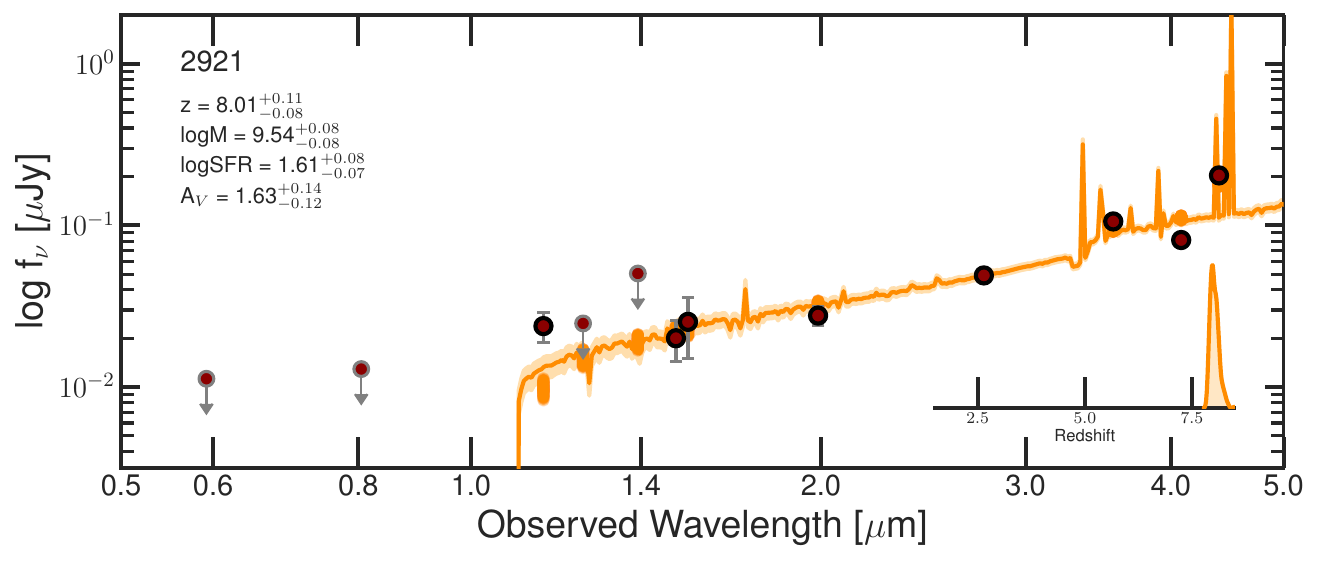}
 \caption{ Examples of four HST-dark galaxies at 
 $\mathrm{z \sim 3.3, \ 3.6, \ 4.6, \ 8}$ 
 that represent the variety of physical properties, isolation/clustering and  morphologies that we find in our sample (see also \citealt{Nelson2022}). 
 (Top panels) Postage stamps in the optical/NIR for four examples of HST-dark galaxies, where the sources show no significant detections in the HST filters. 
 The first row shows HST images with the bands F606W, F814W, F850LP, F125W, F140W, F160W and the Spitzer bands IRAC1, IRAC2 at $\mathrm{3.6 \ and \ 4.5 \mu m }$ respectively. 
  The second row shows new NIRCam JWST images: F115W, F150W, F200W, F277W, F356W, F410M, and F444W. The extraordinary improvement in both depth and resolution of JWST at $3-5\,\mu$m compared to the IRAC images is obvious. Most of these sources would not have been easily selected in previous data and they would have potentially been missing from our cosmic census. 
  (Bottom panels) Posterior spectral energy distributions (yellow; median with 16-84 percentile uncertainties) with the photometry (red dots). Orange points indicate the expected model fluxes from the posterior SEDs. The probability distribution function of the photometric redshift is shown in the lower right part of the panel. In general, HST-dark sources are found to be dust-obscured, massive, star-forming galaxies at $z\sim2-8$.  
 }  
   \label{SEDs+postages}
\end{figure*}

We also include the available ancillary data from HST since the CEERS survey covers the previous CANDELS/EGS field. In particular, we include a re-reduction of all the ACS and WFC3/IR imaging taken in these fields in the five filters: F606W, F814W, F125W, F140W, and F160W. 
The 5$\sigma$ depths measured in the 0\farcs48 circular apertures for the HST data range from 28.4 mag in F606W to 27.0 mag in F140W. While the HST NIR data are shallower, they still provide additional multi-wavelength constraints. Most importantly using the F160W filter allows us to use HST-dark galaxy selections analogous to previous HST+Spitzer searches (see next section).
We match the point-spread functions (PSFs) in all but the WFC3 images to the F444W filter using the deconvolution algorithm from \citet{Lucy1974}. 
The multi-wavelength photometric catalogues were derived with \texttt{SExtractor} \citep[][]{Bertin96}, which we run in dual mode, using a PSF-matched, inverse-variance weighted stack of the F277W, F356W and F444W images as the detection image and measuring fluxes in all bands in circular apertures of 0\farcs48 diameter. 
We scale the obtained fluxes to total based on the AUTO flux measurements in the F444W image with the default Kron parameters. 
In order to take the PSF differences between F444W and the WFC3 filters into account, we match their PSFs to the F160W filter, which has the largest FWHM among all filters considered here. We run \texttt{SExtractor} again on the resulting images, using the same detection image as before. For each WFC3 filter, we then correct the fluxes measured in the original image to match the PSF-matched colour between the WFC3 filter and F444W respectively.
As a general quality cut, the catalogue contains objects with S/N $>$ 5 in at least one of the available NIRCam-wide filters. We also removed objects near the edge and bad pixels, hence we guarantee that all the sources of the catalogue are not spurious.
Finally, to account for systematic uncertainties in the photometry we apply an error floor of 5\% to all flux measurements. 


The final catalogue includes $\mathrm{\sim 26,000}$  
sources with photometry in 12 bands, spanning 0.6 to 4.4 $\mu$m. 

\subsection{HST-dark Galaxy Color Selection with JWST}
\label{final_colourseelection}

Previous HST-dark galaxy studies were based on analyses of HST+Spitzer/IRAC data \citep[e.g.,][]{Wang2016, Wang2019, AlcaldePampliega2019}. These selections were limited by two factors: (1) the very broad PSF of the IRAC data with FWHM of $\sim1.7$ \arcsec\ and (2) the limited depth of available IRAC imaging. Hence, typical HST-dark samples were limited to very bright sources with $\mathrm{[4.5]\lesssim 24.5 mag}$. 
Now, with the unparalleled high-resolution, deep JWST data, we can push such analyses to much fainter sources. Nevertheless, in this first paper, we focus on sources that are identified using very similar selection methods as previous HST+Spitzer samples in order to reveal their nature. 

We use a similar colour cut to the `traditional' HST-dark galaxies selection colour  adopted in the literature \citep[e.g.,][]{Caputi2012,Wang2016}. This colour cut is specifically designed to identify dusty galaxies at $\mathrm{z\sim 3-6}$. Indeed, previous analyses have tentatively confirmed that such red HST-dark sources lie at these redshifts on average. However, the available photometry for these sources is very poor, with significant detections only in the IRAC 3.6 $\mu$m and 4.5 $\mu$m bands, and mostly upper limits at HST wavelengths. 

In this first paper, we substitute the IRAC data for NIRCam data. 
To find an appropriate colour cut, we compute theoretical fluxes using Bayesian Analysis of Galaxies for Physical Inference and Parameter EStimation \citep[BAGPIPES][]{Carnall2018}. We perform similar modelling that we use on the SED fitting (see section \ref{modelingandaverageprop}). We set to a delayed star-formation history (e-folding time $\mathrm{\tau = 3}$ Gyr), a stellar mass of $\mathrm{M_{*}=10\ M_{\odot}}$ and a metallicity of $\mathrm{Z=0.5\ Z_{\odot}}$. We model the fluxes F160W and F444W with a dust attenuation set to $\mathrm{A_{V} = 2.0, \ 3.0, \ 4.0 \ mag}$ respectively tracking the redshift evolution with steeps $\mathrm{\Delta z =0.1}$ between $\mathrm{2<z<6}$. We find that applying a colour cut of $\mathrm{F160W-F444W > 2.0}$ selects galaxies at $\mathrm{z>3}$ at $\mathrm{A_{V}=2.0}$ (see Figure \ref{colour_selection}). 

We thus apply a colour cut of F160W-F444W$>2.0$ to our multi-wavelength catalogue to select HST-dark galaxies. 
After also limiting the sample to $\mathrm{F160W>27 mag}$, we identify a total of 39 galaxies in the $\sim40$ arcmin$^2$ down to an F444W=26.4 magnitude (red diamonds in Figure \ref{colour_selection}). We remove six galaxies that are at the edge of the image and do not have a clear detection in the longer wavelength of the wide NIRCam filters (F444W, F356W, F277W). And three more galaxies for which the photometry is extremely noisy, likely due to remaining detector artefacts. Hence, our final sample consists of 30 HST-dark galaxies. 

\section{SED fitting and variety of HST-dark galaxies.} 
\label{modelingandaverageprop}

We derive the physical properties of the 30 HST-dark galaxies by using Bayesian Analysis of Galaxies for Physical Inference and Parameter EStimation \citep[BAGPIPES][]{Carnall2018}. 
This code generates a complex model
galaxy spectra to fit these to arbitrary combinations of spectroscopic and photometric
data using the MultiNest nested sampling algorithm \citep{Feroz09}. 

The assumptions by \citet{Wang2016, Wang2019} have proven effective to characterize HST-dark galaxies and this work follows the same SED modelling. Specifically, we use a delayed star formation history (SFH) with an e-folding time from $\mathrm{\tau = 0.1-9 Gyr}$ and a uniform prior. Hence, we include relatively short bursts as well as, effectively, a constant SFH.

The stellar models are based on the 2016 updated version of the \citet{Bruzual2003} library with a Kroupa initial mass function. 
We allow for a broad range of metallicities from 0.2 to 2.5 $\mathrm{Z_{\odot}}$. Nebular continuum and line emission are added self-consistently based on the photoionization code CLOUDY \citep{Ferland98} with the ionization parameter set as $\mathrm{log U =-2}$. Finally, we adopt \citet{Calzetti2000} dust attenuation allowing for very heavily attenuated SEDs by setting a range $\mathrm{A_{v} = 0-6\, mag}$, again using a uniform prior. 

The above settings adequately reproduce the range of SEDs of our sample. We find good fits for all our sources. Figure \ref{SEDs+postages} shows postage stamps and SED fits of four of these sources at $\mathrm{z > 3.3}$. They represent the variety of galaxy SEDs that are identified as HST-dark sources, both in terms of physical properties and in terms of morphology. We also show the cutouts of previous IRAC images from the S-CANDELS program \citep{Ashby15}. The outstanding increase in depth and spatial resolution at $3-5\,\mu$m can be appreciated. With the exception of a handful of our sample galaxies, none of these would have been easily selected based on previously available data. 
Hence, the majority of our sample was completely missing from our previous cosmic census of HST+Spitzer. In the following, we use our sample of 30 sources to analyze the physical properties of HST-dark galaxies. The main parameters are summarised in Table \ref{tab:sample} in the Appendix. 

\begin{figure}
 \centering
  \includegraphics[width=\columnwidth]{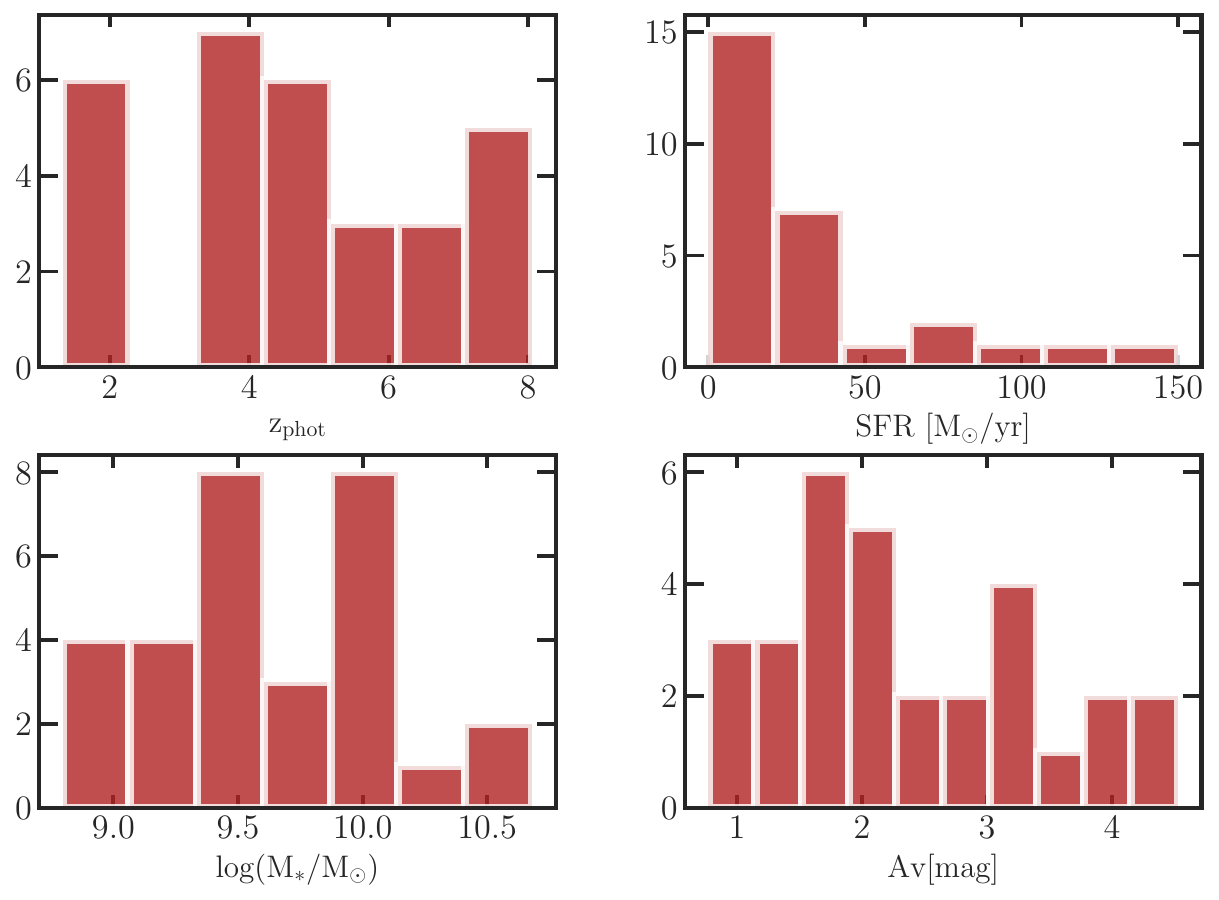}  
 \caption{Histograms of the main physical properties analysed: photometric redshift, SFR (first row), stellar masses and dust attenuation (second row). The histograms show the generally dusty high-z nature of the sample with most of the galaxies at $\mathrm{z > 3.3}$ and a median attenuation of $\mathrm{A_{v} \sim 2 mag}$. The SFRs are moderate, mostly at $\mathrm{SFR < 50 M_{\odot}/year }$ whereas the stellar masses are large with $\mathrm{log(M_{*}/M_{\odot}) \sim 9.5-10}$. Our analysis thus shows that previous stellar mass function estimates at $z>3$ were likely underestimated at high masses.   } 
  \label{histogram}
\end{figure}

\section{HST-dark galaxies: The high-z extension of SMGs? }
\label{results}

\subsection{HST-dark galaxy properties }
Thanks to the very sensitive NIRCam data, we can now -- for the first time -- measure accurate photometric redshifts and SEDs for HST-dark galaxies. Hence, we can now derive significantly more reliable physical parameters such as SFRs and stellar masses compared to previous IRAC-selected HST-dark sources.
In this Section, we thus present the physical properties derived from SED fitting for our final sample of 30 sources (see Figures \ref{SEDs+postages} and \ref{histogram}). 

We find that an HST-dark galaxy selection includes a variety of sources with the vast majority of the sample being high redshift dust-obscured, massive, star-forming systems. The photometric redshifts are accurate for the galaxies at $z > 3.3 $ with small uncertainties ($\Delta z\sim0.1-0.2$) with only one solution in the redshift posterior distribution function (see table \ref{tab:sample} and Figure \ref{SEDs+postages}). The subset of galaxies that is placed at $z < 3.3$ presents more considerable uncertainties and typically shows two possible redshift solutions. 

\begin{figure}
 \centering
      \includegraphics[width=\columnwidth]{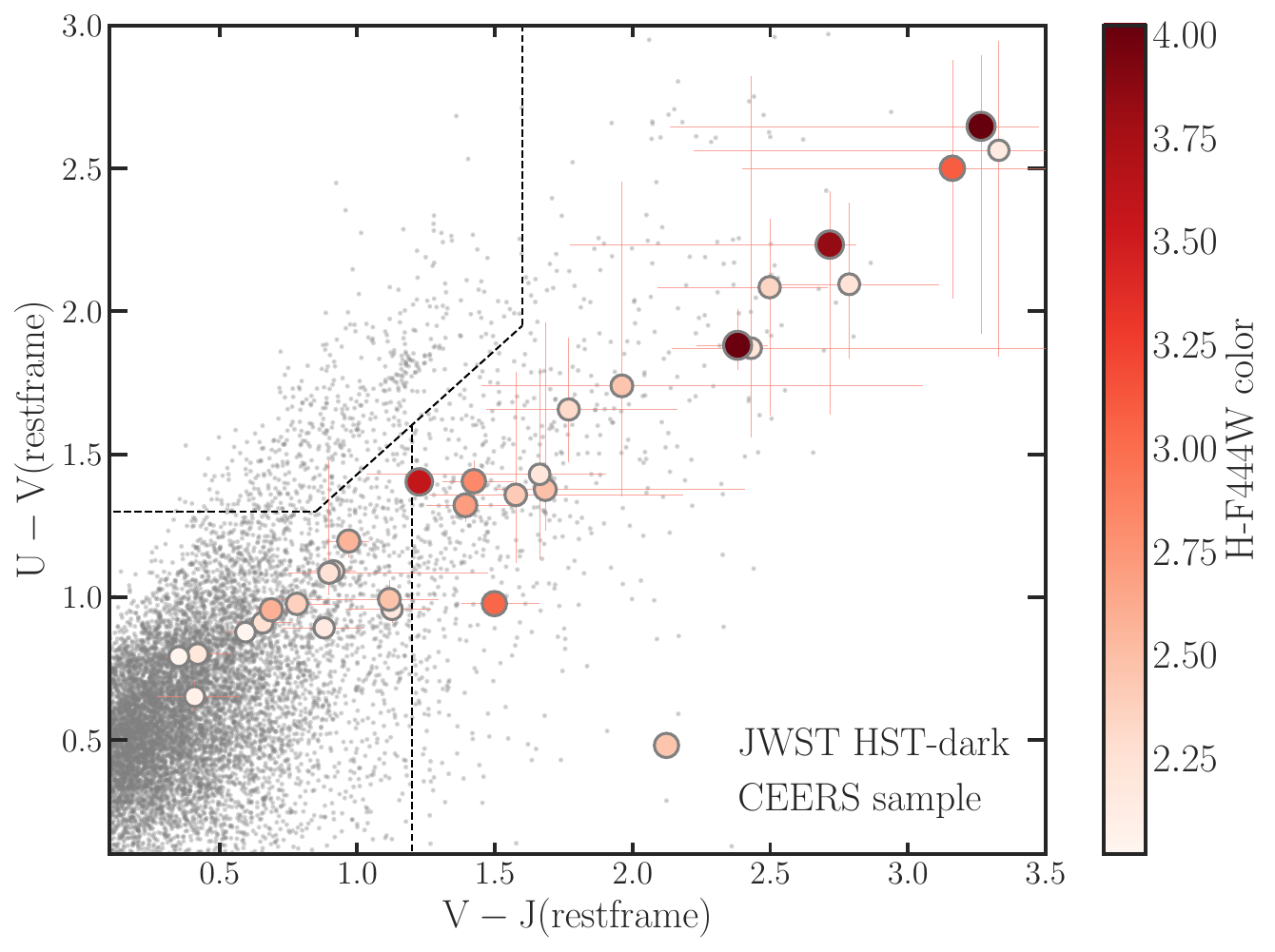}
\caption{ UV against VJ rest-frame colours diagram. HST-dark galaxies are presented as big circles with the H-F444W colour on the third axis whereas the grey dots are the total CEERS sample. The black dashed lines show the usual classification of galaxies in quiescent, star-forming and dusty from \citet{Spitler2014}. Our sample of HST-dark galaxies lies mostly in the dusty colour space with part of the sample in the star-forming galaxies region. There are no quiescent galaxies in our selection due to the colour cut, however, given the photometric uncertainties, we cannot exclude that some of the galaxies are quiescent.} 
  \label{UVJdiagram}
\end{figure}

As expected, we confirm the dusty nature of these colour-selected HST-dark galaxies. We have allowed the dust attenuation to vary broadly in the modelling, but the values only reach up to $\mathrm{A_{v} = 4 mag}$ for almost all HST-dark galaxies with a median value for the sample of $\mathrm{A_{v} \sim 2.1  mag}$. 
An alternative procedure to classify among  dusty and quiescent galaxies is the commonly accepted UVJ diagrams (see e.g. \citealt{Spitler2014}). Figure \ref{UVJdiagram} shows the rest-frame UV colours for the total CEERS sample, corroborating the dusty nature of HST-dark galaxies.

Interestingly, the sample reaches redshifts $\mathrm{z \sim 8}$ for several galaxies. In total, we identify eight 
galaxies at $\mathrm{z>6}$, i.e., in the epoch of reionization, whereas previous HST-dark galaxies selected with Spitzer were typically limited to $z<6$. 
The high attenuation for the galaxies above $\mathrm{z>6}$, with $\mathrm{ A_{V} = 1.7 \pm 0.4}$ suggests the presence of dust in the reionization epoch \citep[see also][]{Fudamoto2021}. We refer to a future paper for more detailed analyses of these sources.  

We have also analysed both the stellar masses and the SFRs from the SED fitting. The SFRs are moderate, when compared to SMGs, with $\mathrm{\sim 80 \%}$ of our sample presenting $\mathrm{SFR < 50 M_{\odot}/yr}$. The stellar masses of our sample are $\mathrm{log(M_{*}/M_{\odot})> 8.8 }$ 
and for the eight galaxies at $\mathrm{z>6}$ we find  $\mathrm{log(M_{*}/M_{\odot})= 9.9 \pm 0.3}$. Hence, we can confirm that HST-dark galaxies are massive with relatively low star formation.

\begin{figure}
 \centering
   \includegraphics[width=0.5\textwidth]{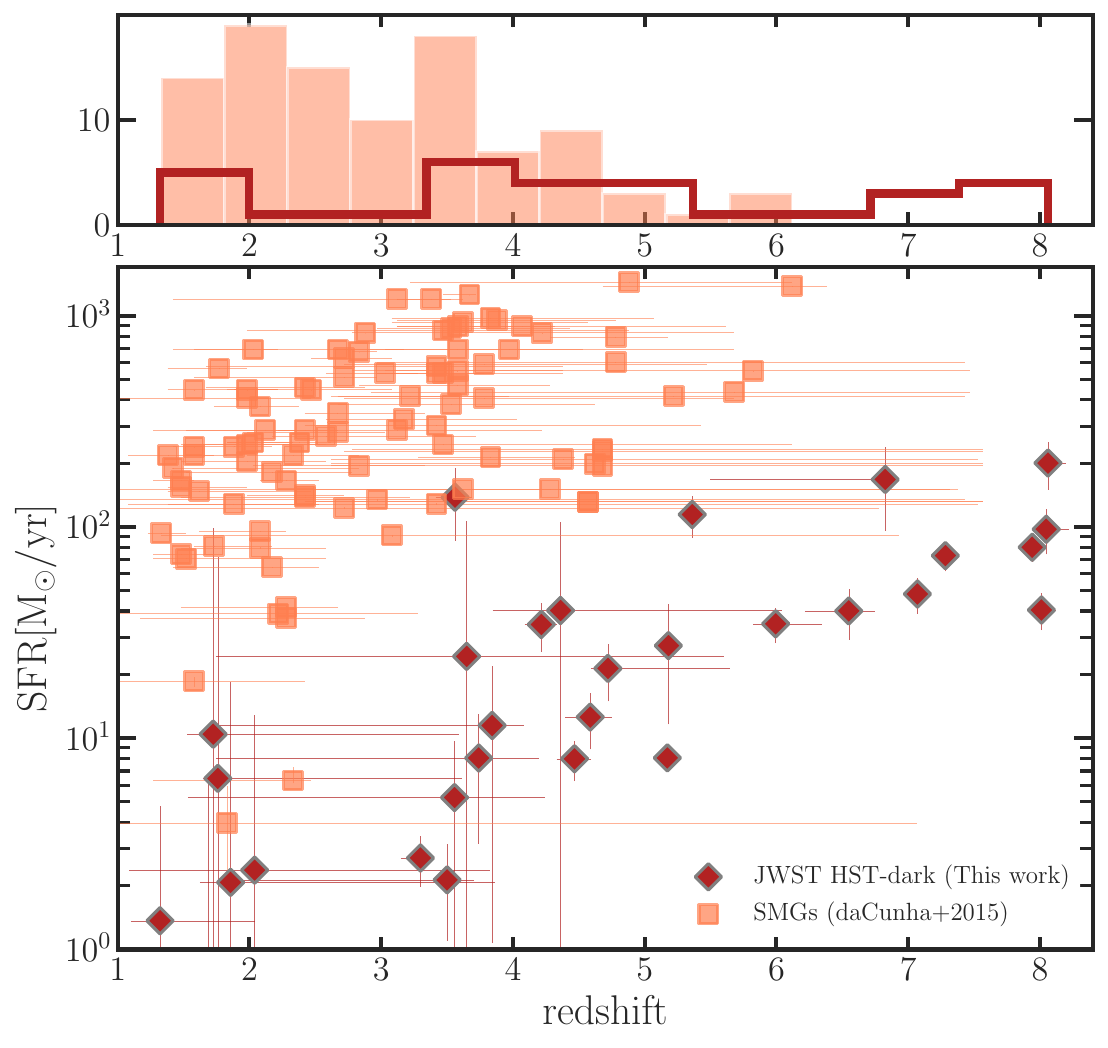}  
   \caption{ HST-dark galaxies compared with submillimetre galaxies (SMGs). (Top panel) The redshift histogram shows that HST-dark galaxies (red line) lie mostly at $\mathrm{ z > 3.2}$ and extend to higher redshift compared to SMGs (orange bars).  
   (Bottom panel) Photometric redshifts against star formation rates. On average, HST-dark galaxies lie at higher redshift and have significantly lower SFRs  (orange squares; ALESS sample from \citet{daCunha2015}). The significantly smaller error bars in the photometric redshifts are due to the improvement coming from the deep, high-resolution JWST/NIRCam data. 
     } 
  \label{SFR_z_TOT_ALESS}
\end{figure}

\subsection{HST-dark galaxies compared to SMGs. } 

Given their dusty nature, it is interesting to compare the HST-dark galaxies with a classical SMG sample, in order to put them in a broader context of massive high-z dusty galaxies. We find that HST-dark galaxies have higher redshifts compared to SMGs (see the histogram in Figure  \ref{SFR_z_TOT_ALESS}).  
Interestingly and probably not unexpectedly, HST-dark galaxies have lower SFRs than SMGs. Similarly, we also find HST-dark galaxies have more than an order of magnitude lower stellar masses than the massive SMGs that weigh $\mathrm{M_{*} \sim 10^{12} M_{\odot}}$. 

\begin{figure}
 \centering
   \includegraphics[width=0.5\textwidth]{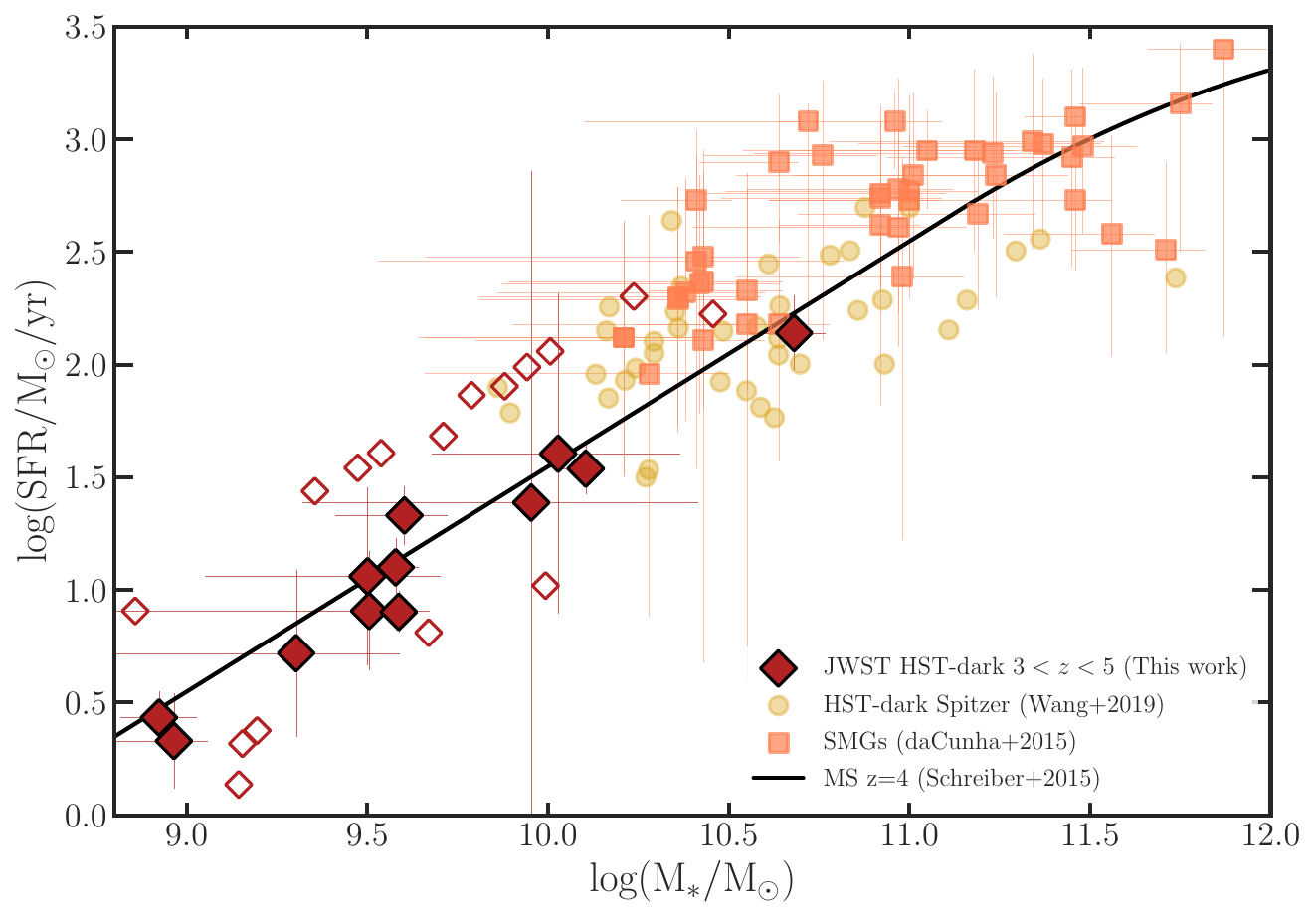}  \caption{Star formation rate against stellar mass for our sample corresponding to $3 < z < 5$ (red diamonds) and the total sample (empty diamonds). The black line shows the MS at $z = 4$ from \citet{Schreiber2015}. 
   HST-dark galaxies lie on the MS as other optically dark massive galaxies selected with Spitzer data from \citet{Wang2019} (yellow dots). SMGs from \citet{daCunha2015} are above the MS (orange squares). } 
     \label{MS}
\end{figure}

In Figure \ref{MS}, we further show the location of HST-dark and SMGs on the SFR-$\mathrm{M_{*}}$ diagram with respect to the main sequence (MS) of galaxies at $ \mathrm{z\sim4}$. As can be seen, HST-dark sources at these redshifts follow the MS. They populate its lower mass end compared to both SMGs and IRAC-selected HST-dark galaxies. 
Specifically, we have compared our results with two previous studies of dusty galaxies: The ALESS sample from \citet{daCunha2015} shows a larger number of starburst galaxies that lie significantly above the MS. In between our sample and SMGs, we find more massive HST-dark galaxies from \citet{Wang2019}. This difference could be due to the selection criteria and lower mass HST-dark galaxies could have been missing from Spitzer-selected samples. 

In a nutshell, HST-dark galaxies appear to be the higher redshift, lower SFR extension of SMGs. They have likely just not been detected in previous sub-millimetre single-dish studies due to their faintness. The synergy between JWST and ALMA data will be needed to confirm this rather interesting outcome.

\section{Cosmic Star Formation Rate: is HST missing a significant contribution at high-z?}
\label{SFRDsection}

The cosmic star formation history is key to explaining the mass assembly and the metal enrichment of the galaxy population 
both vital to understanding galaxy evolution. Although the Cosmic Star Formation Rate Density (CSFRD) is well constrained up to $\mathrm{z \sim 8}$ for UV-selected galaxies (i.e., LBGs), the contribution from obscured, IR rest-frame selected galaxies is still highly uncertain and only poorly constrained at $\mathrm{z>3}$ \citep[e.g.,][]{Casey2014, Casey2019, Zavala2021}. A central question is: when does the dust-obscured star formation start to dominate the CSFRD? And, more specifically, what is the contribution to the CSFRD from HST-dark galaxies? In this Section, we approach this question and put our results in the context of the most relevant high-$z$ galaxy populations. 

We calculate the SFRD for our sample of HST-dark galaxies at $\mathrm{z>3.5}$ in four 
redshift equidistant bins 
centred at $z = $ 3.3, 4.5, 5.7 and 6.9 
respectively. 
We find a contribution of $\mathrm{SFRD = 1.4^{+1.1}_{-0.8} \times 10^{-3} M_{\odot}/yr/Mpc^{3}}$ at $z = 4.5$ that remains constant within the uncertainty up to $z\sim7$. Specifically, we measure $\mathrm{SFRD \sim 2.0^{+2.0}_{-1.1} \times 10^{-3} M_{\odot}/yr/Mpc^{3}}$ at $\mathrm{z \sim 6}$, and $\mathrm{SFRD \sim 3.2^{+1.8}_{-1.3} \times 10^{-3} M_{\odot}/yr/Mpc^{3}}$ at $z\sim7$. 

\begin{figure*}
 \centering
      \includegraphics[width=2\columnwidth]{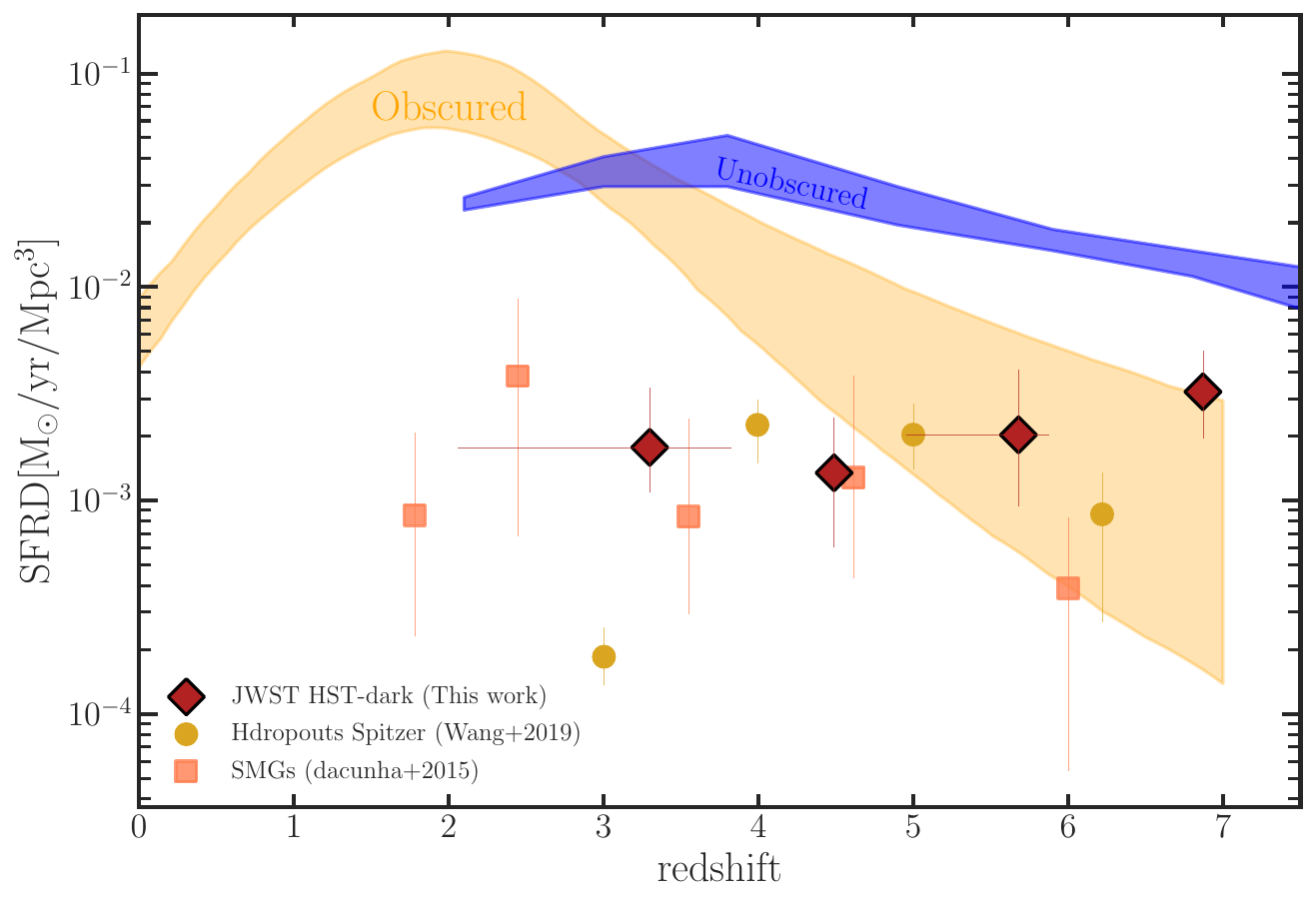}
\caption{ Star formation rate density against redshift for the HST-dark galaxies (red diamonds) compared to a compilation of previous studies. The orange area represents the dust-obscured star formation history from semiempirical models presented in \citet{Zavala2021} whereas the blue area shows the unobscured star formation \citep{Bouwens2022}. 
The contribution of our HST-dark galaxies remains almost constant from redshift $3.5 < z < 6$. 
The HST-dark contribution is similar to SMGs at $z \sim 5-6$ (orange squares; \citealt{daCunha2015}). 
Our SFRD is in agreement with previous HST-dark studies at $z \sim 4-6 $  (orange dots; \citealt{Wang2019}) but we extend the result to $z \sim 7$. 
 } 
  \label{SFRD}
\end{figure*}

Figure \ref{SFRD} shows that our results have similar SFRD than previously selected HST-dark galaxy samples at  $z \sim 4-6 $. There are no current observations of HST-dark galaxies at $\mathrm{z>6}$. Our results however suggest that the SFRD of HST-dark galaxies remains constant, within the uncertainty, or becomes even slightly larger at $z \sim 7$. This could suggest a remarkable presence of dust in the epoch of reionization. As a caveat, we notice that our selection criteria tend to select HST-dark galaxies at $\mathrm{z \sim 7}$ due to the bright F444W flux that selects [OIII] lines in the SED. Also, our redshift bins currently only contain a few sources 
each and future studies with larger samples are needed to corroborate this result. 

Our analysis shows an HST-dark contribution to the SFRD similar to SMGs at $\mathrm{4 < z < 5}$, but larger at $\mathrm{z>5}$. An evident explanation is that the number of SMGs drops at $\mathrm{z>5}$ \citep{daCunha2015}. Even if their SFRs are an order of magnitude higher than the average of HST-dark galaxies, their SFRD is lower at high-$z$.  

Our study shows that HST-dark galaxies  contribute an order of magnitude less than 2mm-selected DSFGs at $\mathrm{z \sim 4}$ from the MORA survey presented in \citet{Casey2021}. The reason probably is the lower SFRs and this suggests that fainter dusty galaxies could have been missed by shallow ALMA surveys at $\mathrm{z \sim4-7}$. 

Our study, and previous studies mentioned above, indicate that the dusty population at high-$z$ is larger than what was expected. 
In particular, more extensive samples are needed from deeper surveys and multi-wavelength observations to understand how the disparate objects seen today relate and contribute to the overall cosmic SFRD. A very important result from our analysis is that HST-dark galaxies significantly contribute to the CSFRD, in particular at $\mathrm{z > 5}$. 

\section{Summary and Conclusions} 
\label{Summary}

In this work, we have exploited some of the first JWST/NIRCam data to identify red high-$z$ galaxies that were missing from rest-frame UV selections based on HST. These HST-dark galaxies were selected based on $\mathrm{H-F444W}$ colours. We studied 30 galaxies in the four ERS CEERS NIRCam pointings and used SED fitting to derive their properties and put them into context by comparison to other dusty high-redshift galaxy populations. 

The main results are summarized as follows:

\begin{itemize}

 \item The average properties of HST-dark galaxies show their high-$z$ dusty nature with $\mathrm{80 \ \%}$ of the sample at $3 \lesssim z \lesssim 8$. Surprisingly, we identify eighth galaxies at $z>6$ with a median dust attenuation of $\mathrm{A_{v} \sim 1.9 \pm 0.4}$ mag. This indicates that massive, heavily obscured galaxies have been missing from our cosmic census in the epoch of reionization.

 \item We find that HST-dark galaxies have moderate star formation rates compared to SMGs with $\mathrm{ SFR < 50\,M_{\odot}/yr }$, but they extend to higher redshifts. HST-dark galaxies lie on the MS of galaxies with lower masses than SMGs, but still with a median of $\mathrm{log(M_{*}/M_{\odot})= 9.9 \pm 0.3}$ for the eight galaxies at $z>6$. 

 \item We find that the contribution of HST-dark galaxies to the cosmic star formation rate density (CSFRD) is 
 $\mathrm{SFRD \sim 3.2^{+1.8}_{-1.3} \times 10^{-3} M_{\odot}/yr/Mpc^{3}}$ at $z\sim7$ at $\mathrm{z \sim 7}$ remaining almost constant over redshift $\mathrm{3.3 < z < 7}$. Our SFRD results are in broad agreement with previous HST-dark galaxy studies, but the JWST-selected sample indicates that HST-dark galaxies existed up to higher redshifts than previously assumed, i.e. $\mathrm{z \sim 8}$.

\end{itemize}

Thanks to the outstanding improvement of NIRCam data over the previous Spitzer/IRAC imaging, this work provides  more reliable photometric redshifts, SFRs, and stellar masses compared to previous HST-dark galaxy studies. Hence, we can finally conclude that HST-dark galaxies have a high-z nature and that they are numerous at $z >3$, with a surface density of $\sim0.8$ arcmin$^{-2}$. The high number of dusty galaxies with $\mathrm{z > 6}$ suggests that the presence of dust is relevant in the reionization epoch. 

The SFRD contribution of our sample of HST-dark galaxies is in agreement with previous H-dropout studies at  $z<6$. We also find a significant SFRD at $z \sim 7$ that needs to be corroborated with larger samples of HST-dark galaxies. But already now, with these first NIRCam images, JWST has demonstrated its power to identify more complete samples of rest-frame optically selected galaxies out to very high redshifts. We have shown that dusty galaxies at $z>3$ are likely to be far more numerous than expected and that such HST-dark galaxies provide an important, but hitherto underestimated contribution to the high mass end of the galaxy mass function and the cosmic SFRD up to $z\sim7-8$.

\section*{Acknowledgements}

We acknowledge support from the Swiss National Science Foundation through project grant 200020\_207349 (LB, PAO, and AW). We also acknowledge the anonymous referee for the constructive feedback which resulted in an improved version of this manuscript. 
The Cosmic Dawn Center (DAWN) is funded by the Danish National Research Foundation under grant No.\ 140. 
YF acknowledge support from NAOJ ALMA Scientific Research Grant number 2020-16B. 
VG acknowledges support by the ANID BASAL projects ACE210002 and FB210003
KEH acknowledges support from the Carlsberg Foundation Reintegration Fellowship Grant CF21-0103. 
RS acknowledges an STFC Ernest Rutherford Fellowship (ST/S004831/1)
MS acknowledges support from the CIDEGENT/2021/059 grant, and from project PID2019-109592GB-I00/AEI/10.13039/501100011033 from the Spanish Ministerio de Ciencia e Innovacion - Agencia Estatal de Investigacion.
RJB and MS acknowledge support from NWO grant TOP1.16.057.

Cloud-based data processing and file storage for this work is provided by the AWS Cloud Credits for Research program.

This work is based on observations made with the NASA/ESA/CSA James Webb Space Telescope. The data were obtained from the Mikulski Archive for Space Telescopes at the Space Telescope Science Institute, which is operated by the Association of Universities for Research in Astronomy, Inc., under NASA contract NAS 5-03127 for JWST. These observations are associated with program \# 1345.

Facilities: \textit{JWST}, \textit{HST}

Software:
    \texttt{matplotlib} \citep{matplotlib},
    \texttt{numpy} \citep{numpy},
    \texttt{scipy} \citep{scipy},
    \texttt{jupyter} \citep{jupyter},
    \texttt{Astropy}
    \citep{astropy1, astropy2},
    \texttt{grizli} \citep{grizli},
    \texttt{SExtractor} \citep{Bertin96},
    \texttt{bagpipes} \citep{Carnall2018}

\section*{Data Availability}

The calibrated (stage 2) data used here are available Mikulski Archive for Space Telescopes (\url{https://mast.stsci.edu}). Further data products are available from the authors upon reasonable request.



\bibliographystyle{mnras}
\bibliography{Hdropspaper} 

\begin{thebibliography}{}
\makeatletter
\relax
\def\mn@urlcharsother{\let\do\@makeother \do\$\do\&\do\#\do\^\do\_\do\%\do\~}
\def\mn@doi{\begingroup\mn@urlcharsother \@ifnextchar [ {\mn@doi@}
  {\mn@doi@[]}}
\def\mn@doi@[#1]#2{\def\@tempa{#1}\ifx\@tempa\@empty \href
  {http://dx.doi.org/#2} {doi:#2}\else \href {http://dx.doi.org/#2} {#1}\fi
  \endgroup}
\def\mn@eprint#1#2{\mn@eprint@#1:#2::\@nil}
\def\mn@eprint@arXiv#1{\href {http://arxiv.org/abs/#1} {{\tt arXiv:#1}}}
\def\mn@eprint@dblp#1{\href {http://dblp.uni-trier.de/rec/bibtex/#1.xml}
  {dblp:#1}}
\def\mn@eprint@#1:#2:#3:#4\@nil{\def\@tempa {#1}\def\@tempb {#2}\def\@tempc
  {#3}\ifx \@tempc \@empty \let \@tempc \@tempb \let \@tempb \@tempa \fi \ifx
  \@tempb \@empty \def\@tempb {arXiv}\fi \@ifundefined
  {mn@eprint@\@tempb}{\@tempb:\@tempc}{\expandafter \expandafter \csname
  mn@eprint@\@tempb\endcsname \expandafter{\@tempc}}}

\bibitem[\protect\citeauthoryear{{Alcalde Pampliega} et~al.,}{{Alcalde
  Pampliega} et~al.}{2019}]{AlcaldePampliega2019}
{Alcalde Pampliega} B.,  et~al., 2019, \mn@doi [\apj]
  {10.3847/1538-4357/ab14f2}, \href
  {https://ui.adsabs.harvard.edu/abs/2019ApJ...876..135A} {876, 135}

\bibitem[\protect\citeauthoryear{{Ashby} et~al.,}{{Ashby}
  et~al.}{2015}]{Ashby15}
{Ashby} M.~L.~N.,  et~al., 2015, \mn@doi [\apjs] {10.1088/0067-0049/218/2/33},
  \href {http://adsabs.harvard.edu/abs/2015ApJS..218...33A} {218, 33}

\bibitem[\protect\citeauthoryear{{Astropy Collaboration} et~al.,}{{Astropy
  Collaboration} et~al.}{2013}]{astropy1}
{Astropy Collaboration} et~al., 2013, \mn@doi [\aap]
  {10.1051/0004-6361/201322068}, \href
  {http://adsabs.harvard.edu/abs/2013A%26A...558A..33A} {558, A33}

\bibitem[\protect\citeauthoryear{{Astropy Collaboration} et~al.,}{{Astropy
  Collaboration} et~al.}{2018}]{astropy2}
{Astropy Collaboration} et~al., 2018, \mn@doi [\aj] {10.3847/1538-3881/aabc4f},
  \href {https://ui.adsabs.harvard.edu/abs/2018AJ....156..123A} {156, 123}

\bibitem[\protect\citeauthoryear{{Bertin} \& {Arnouts}}{{Bertin} \&
  {Arnouts}}{1996}]{Bertin96}
{Bertin} E.,  {Arnouts} S.,  1996, \aaps, \href
  {http://adsabs.harvard.edu/abs/1996A%26AS..117..393B} {117, 393}

\bibitem[\protect\citeauthoryear{{Bouwens}, {Illingworth}, {Ellis}, {Oesch}  \&
  {Stefanon}}{{Bouwens} et~al.}{2022}]{Bouwens2022}
{Bouwens} R.~J.,  {Illingworth} G.~D.,  {Ellis} R.~S.,  {Oesch} P.~A.,
  {Stefanon} M.,  2022, arXiv e-prints, \href
  {https://ui.adsabs.harvard.edu/abs/2022arXiv220511526B} {p. arXiv:2205.11526}

\bibitem[\protect\citeauthoryear{{Brammer}}{{Brammer}}{2018}]{grizli}
{Brammer} G.,  2018, {Gbrammer/Grizli: Preliminary Release}, Zenodo,
  \mn@doi{10.5281/zenodo.1146905}

\bibitem[\protect\citeauthoryear{{Bruzual} \& {Charlot}}{{Bruzual} \&
  {Charlot}}{2003}]{Bruzual2003}
{Bruzual} G.,  {Charlot} S.,  2003, \mn@doi [\mnras]
  {10.1046/j.1365-8711.2003.06897.x}, \href
  {http://adsabs.harvard.edu/abs/2003MNRAS.344.1000B} {344, 1000}

\bibitem[\protect\citeauthoryear{{Calzetti}, {Armus}, {Bohlin}, {Kinney},
  {Koornneef}  \& {Storchi-Bergmann}}{{Calzetti} et~al.}{2000}]{Calzetti2000}
{Calzetti} D.,  {Armus} L.,  {Bohlin} R.~C.,  {Kinney} A.~L.,  {Koornneef} J.,
   {Storchi-Bergmann} T.,  2000, \mn@doi [\apj] {10.1086/308692}, \href
  {http://adsabs.harvard.edu/abs/2000ApJ...533..682C} {533, 682}

\bibitem[\protect\citeauthoryear{{Caputi} et~al.,}{{Caputi}
  et~al.}{2012}]{Caputi2012}
{Caputi} K.~I.,  et~al., 2012, \mn@doi [\apjl] {10.1088/2041-8205/750/1/L20},
  \href {https://ui.adsabs.harvard.edu/abs/2012ApJ...750L..20C} {750, L20}

\bibitem[\protect\citeauthoryear{{Caputi} et~al.,}{{Caputi}
  et~al.}{2015}]{Caputi2015}
{Caputi} K.~I.,  et~al., 2015, \mn@doi [\apj] {10.1088/0004-637X/810/1/73},
  \href {https://ui.adsabs.harvard.edu/abs/2015ApJ...810...73C} {810, 73}

\bibitem[\protect\citeauthoryear{{Carnall}, {McLure}, {Dunlop}  \&
  {Dav{\'e}}}{{Carnall} et~al.}{2018}]{Carnall2018}
{Carnall} A.~C.,  {McLure} R.~J.,  {Dunlop} J.~S.,   {Dav{\'e}} R.,  2018,
  \mn@doi [\mnras] {10.1093/mnras/sty2169}, \href
  {https://ui.adsabs.harvard.edu/abs/2018MNRAS.480.4379C} {480, 4379}

\bibitem[\protect\citeauthoryear{{Carnall} et~al.,}{{Carnall}
  et~al.}{2020}]{Carnall2020}
{Carnall} A.~C.,  et~al., 2020, \mn@doi [\mnras] {10.1093/mnras/staa1535},
  \href {https://ui.adsabs.harvard.edu/abs/2020MNRAS.496..695C} {496, 695}

\bibitem[\protect\citeauthoryear{{Casey}, {Narayanan}  \& {Cooray}}{{Casey}
  et~al.}{2014}]{Casey2014}
{Casey} C.~M.,  {Narayanan} D.,   {Cooray} A.,  2014, \mn@doi [\physrep]
  {10.1016/j.physrep.2014.02.009}, \href
  {https://ui.adsabs.harvard.edu/abs/2014PhR...541...45C} {541, 45}

\bibitem[\protect\citeauthoryear{{Casey} et~al.,}{{Casey}
  et~al.}{2018}]{Casey2018}
{Casey} C.~M.,  et~al., 2018, \mn@doi [\apj] {10.3847/1538-4357/aac82d}, \href
  {https://ui.adsabs.harvard.edu/abs/2018ApJ...862...77C} {862, 77}

\bibitem[\protect\citeauthoryear{{Casey} et~al.,}{{Casey}
  et~al.}{2019}]{Casey2019}
{Casey} C.,  et~al., 2019, \baas, \href
  {https://ui.adsabs.harvard.edu/abs/2019BAAS...51c.212C} {51, 212}

\bibitem[\protect\citeauthoryear{{Casey} et~al.,}{{Casey}
  et~al.}{2021}]{Casey2021}
{Casey} C.~M.,  et~al., 2021, \mn@doi [\apj] {10.3847/1538-4357/ac2eb4}, \href
  {https://ui.adsabs.harvard.edu/abs/2021ApJ...923..215C} {923, 215}

\bibitem[\protect\citeauthoryear{{Ferland}, {Korista}, {Verner}, {Ferguson},
  {Kingdon}  \& {Verner}}{{Ferland} et~al.}{1998}]{Ferland98}
{Ferland} G.~J.,  {Korista} K.~T.,  {Verner} D.~A.,  {Ferguson} J.~W.,
  {Kingdon} J.~B.,   {Verner} E.~M.,  1998, \mn@doi [\pasp] {10.1086/316190},
  \href {http://adsabs.harvard.edu/abs/1998PASP..110..761F} {110, 761}

\bibitem[\protect\citeauthoryear{{Feroz}, {Hobson}  \& {Bridges}}{{Feroz}
  et~al.}{2009}]{Feroz09}
{Feroz} F.,  {Hobson} M.~P.,   {Bridges} M.,  2009, \mn@doi [\mnras]
  {10.1111/j.1365-2966.2009.14548.x}, \href
  {https://ui.adsabs.harvard.edu/abs/2009MNRAS.398.1601F} {398, 1601}

\bibitem[\protect\citeauthoryear{{Finkelstein} et~al.,}{{Finkelstein}
  et~al.}{2022}]{Finkelstein2022}
{Finkelstein} S.~L.,  et~al., 2022, arXiv e-prints, \href
  {https://ui.adsabs.harvard.edu/abs/2022arXiv221105792F} {p. arXiv:2211.05792}

\bibitem[\protect\citeauthoryear{{Franco} et~al.,}{{Franco}
  et~al.}{2018}]{Franco2018}
{Franco} M.,  et~al., 2018, \mn@doi [\aap] {10.1051/0004-6361/201832928}, \href
  {https://ui.adsabs.harvard.edu/abs/2018A&A...620A.152F} {620, A152}

\bibitem[\protect\citeauthoryear{{Fudamoto} et~al.,}{{Fudamoto}
  et~al.}{2021}]{Fudamoto2021}
{Fudamoto} Y.,  et~al., 2021, \mn@doi [\nat] {10.1038/s41586-021-03846-z},
  \href {https://ui.adsabs.harvard.edu/abs/2021Natur.597..489F} {597, 489}

\bibitem[\protect\citeauthoryear{{Gruppioni} et~al.,}{{Gruppioni}
  et~al.}{2020}]{Gruppioni2020}
{Gruppioni} C.,  et~al., 2020, \mn@doi [\aap] {10.1051/0004-6361/202038487},
  \href {https://ui.adsabs.harvard.edu/abs/2020A&A...643A...8G} {643, A8}

\bibitem[\protect\citeauthoryear{{Huang}, {Zheng}, {Rigopoulou}, {Magdis},
  {Fazio}  \& {Wang}}{{Huang} et~al.}{2011}]{Huang2011}
{Huang} J.~S.,  {Zheng} X.~Z.,  {Rigopoulou} D.,  {Magdis} G.,  {Fazio} G.~G.,
   {Wang} T.,  2011, \mn@doi [\apjl] {10.1088/2041-8205/742/1/L13}, \href
  {https://ui.adsabs.harvard.edu/abs/2011ApJ...742L..13H} {742, L13}

\bibitem[\protect\citeauthoryear{Hunter}{Hunter}{2007}]{matplotlib}
Hunter J.~D.,  2007, \mn@doi [Computing In Science \& Engineering]
  {10.1109/MCSE.2007.55}, 9, 90

\bibitem[\protect\citeauthoryear{Kluyver et~al.,}{Kluyver
  et~al.}{2016}]{jupyter}
Kluyver T.,  et~al., 2016, in Loizides F.,  Schmidt B.,  eds, Positioning and
  Power in Academic Publishing: Players, Agents and Agendas. pp 87 -- 90

\bibitem[\protect\citeauthoryear{{Labbe} et~al.,}{{Labbe}
  et~al.}{2022}]{Labbe22}
{Labbe} I.,  et~al., 2022, arXiv e-prints, \href
  {https://ui.adsabs.harvard.edu/abs/2022arXiv220712446L} {p. arXiv:2207.12446}

\bibitem[\protect\citeauthoryear{{Lucy}}{{Lucy}}{1974}]{Lucy1974}
{Lucy} L.~B.,  1974, \mn@doi [\aj] {10.1086/111605}, \href
  {https://ui.adsabs.harvard.edu/abs/1974AJ.....79..745L} {79, 745}

\bibitem[\protect\citeauthoryear{{Manning} et~al.,}{{Manning}
  et~al.}{2022}]{Manning2022}
{Manning} S.~M.,  et~al., 2022, \mn@doi [\apj] {10.3847/1538-4357/ac366a},
  \href {https://ui.adsabs.harvard.edu/abs/2022ApJ...925...23M} {925, 23}

\bibitem[\protect\citeauthoryear{{Marrone} et~al.,}{{Marrone}
  et~al.}{2018}]{Marrone2018}
{Marrone} D.~P.,  et~al., 2018, \mn@doi [\nat] {10.1038/nature24629}, \href
  {https://ui.adsabs.harvard.edu/abs/2018Natur.553...51M} {553, 51}

\bibitem[\protect\citeauthoryear{{Naidu} et~al.,}{{Naidu}
  et~al.}{2022}]{Naidu22}
{Naidu} R.~P.,  et~al., 2022, arXiv e-prints, \href
  {https://ui.adsabs.harvard.edu/abs/2022arXiv220709434N} {p. arXiv:2207.09434}

\bibitem[\protect\citeauthoryear{{Nelson} et~al.,}{{Nelson}
  et~al.}{2022}]{Nelson2022}
{Nelson} E.~J.,  et~al., 2022, arXiv e-prints, \href
  {https://ui.adsabs.harvard.edu/abs/2022arXiv220801630N} {p. arXiv:2208.01630}

\bibitem[\protect\citeauthoryear{{Oke} \& {Gunn}}{{Oke} \&
  {Gunn}}{1983}]{Oke83}
{Oke} J.~B.,  {Gunn} J.~E.,  1983, \mn@doi [\apj] {10.1086/160817}, \href
  {http://adsabs.harvard.edu/abs/1983ApJ...266..713O} {266, 713}

\bibitem[\protect\citeauthoryear{Oliphant}{Oliphant}{2015}]{numpy}
Oliphant T.~E.,  2015, Guide to NumPy.
Continuum Press

\bibitem[\protect\citeauthoryear{{Planck Collaboration} et~al.,}{{Planck
  Collaboration} et~al.}{2016}]{Planck2015}
{Planck Collaboration} et~al., 2016, \mn@doi [\aap]
  {10.1051/0004-6361/201525830}, \href
  {https://ui.adsabs.harvard.edu/abs/2016A&A...594A..13P} {594, A13}

\bibitem[\protect\citeauthoryear{{Riechers} et~al.,}{{Riechers}
  et~al.}{2013}]{Riechers2013}
{Riechers} D.~A.,  et~al., 2013, \mn@doi [\nat] {10.1038/nature12050}, \href
  {http://adsabs.harvard.edu/abs/2013Natur.496..329R} {496, 329}

\bibitem[\protect\citeauthoryear{{Santini} et~al.,}{{Santini}
  et~al.}{2020}]{Santini2021}
{Santini} P.,  et~al., 2020, arXiv e-prints, \href
  {https://ui.adsabs.harvard.edu/abs/2020arXiv201110584S} {p. arXiv:2011.10584}

\bibitem[\protect\citeauthoryear{{Santini} et~al.,}{{Santini}
  et~al.}{2022}]{Santini22}
{Santini} P.,  et~al., 2022, arXiv e-prints, \href
  {https://ui.adsabs.harvard.edu/abs/2022arXiv220711379S} {p. arXiv:2207.11379}

\bibitem[\protect\citeauthoryear{{Schreiber} et~al.,}{{Schreiber}
  et~al.}{2015}]{Schreiber2015}
{Schreiber} C.,  et~al., 2015, \mn@doi [\aap] {10.1051/0004-6361/201425017},
  \href {https://ui.adsabs.harvard.edu/abs/2015A&A...575A..74S} {575, A74}

\bibitem[\protect\citeauthoryear{{Simpson} et~al.,}{{Simpson}
  et~al.}{2014}]{Simpson2014}
{Simpson} J.~M.,  et~al., 2014, \mn@doi [\apj] {10.1088/0004-637X/788/2/125},
  \href {https://ui.adsabs.harvard.edu/abs/2014ApJ...788..125S} {788, 125}

\bibitem[\protect\citeauthoryear{{Spitler} et~al.,}{{Spitler}
  et~al.}{2014}]{Spitler2014}
{Spitler} L.~R.,  et~al., 2014, \mn@doi [\apjl] {10.1088/2041-8205/787/2/L36},
  \href {https://ui.adsabs.harvard.edu/abs/2014ApJ...787L..36S} {787, L36}

\bibitem[\protect\citeauthoryear{{Stefanon} et~al.,}{{Stefanon}
  et~al.}{2015}]{Stefanon2015}
{Stefanon} M.,  et~al., 2015, \mn@doi [\apj] {10.1088/0004-637X/803/1/11},
  \href {https://ui.adsabs.harvard.edu/abs/2015ApJ...803...11S} {803, 11}

\bibitem[\protect\citeauthoryear{{Sun} et~al.,}{{Sun} et~al.}{2021}]{Sun2021}
{Sun} F.,  et~al., 2021, arXiv e-prints, \href
  {https://ui.adsabs.harvard.edu/abs/2021arXiv210901751S} {p. arXiv:2109.01751}

\bibitem[\protect\citeauthoryear{{Tanaka} et~al.,}{{Tanaka}
  et~al.}{2019}]{Tanaka2019}
{Tanaka} M.,  et~al., 2019, \mn@doi [\apjl] {10.3847/2041-8213/ab4ff3}, \href
  {https://ui.adsabs.harvard.edu/abs/2019ApJ...885L..34T} {885, L34}

\bibitem[\protect\citeauthoryear{{Umehata} et~al.,}{{Umehata}
  et~al.}{2020}]{Umehata2020}
{Umehata} H.,  et~al., 2020, \mn@doi [\aap] {10.1051/0004-6361/202038146},
  \href {https://ui.adsabs.harvard.edu/abs/2020A&A...640L...8U} {640, L8}

\bibitem[\protect\citeauthoryear{{Valentino} et~al.,}{{Valentino}
  et~al.}{2020}]{Valentino2020}
{Valentino} F.,  et~al., 2020, \mn@doi [\apj] {10.3847/1538-4357/ab64dc}, \href
  {https://ui.adsabs.harvard.edu/abs/2020ApJ...889...93V} {889, 93}

\bibitem[\protect\citeauthoryear{{Virtanen} et~al.,}{{Virtanen}
  et~al.}{2020}]{scipy}
{Virtanen} P.,  et~al., 2020, \mn@doi [Nature Methods]
  {10.1038/s41592-019-0686-2}, \href
  {https://ui.adsabs.harvard.edu/abs/2020NatMe..17..261V} {17, 261}

\bibitem[\protect\citeauthoryear{{Wang} et~al.,}{{Wang}
  et~al.}{2016}]{Wang2016}
{Wang} T.,  et~al., 2016, \mn@doi [\apj] {10.3847/0004-637X/816/2/84}, \href
  {https://ui.adsabs.harvard.edu/abs/2016ApJ...816...84W} {816, 84}

\bibitem[\protect\citeauthoryear{{Wang} et~al.,}{{Wang}
  et~al.}{2019}]{Wang2019}
{Wang} T.,  et~al., 2019, \mn@doi [\nat] {10.1038/s41586-019-1452-4}, \href
  {https://ui.adsabs.harvard.edu/abs/2019Natur.572..211W} {572, 211}

\bibitem[\protect\citeauthoryear{{Williams} et~al.,}{{Williams}
  et~al.}{2019}]{Williams2019}
{Williams} C.~C.,  et~al., 2019, \mn@doi [\apj] {10.3847/1538-4357/ab44aa},
  \href {https://ui.adsabs.harvard.edu/abs/2019ApJ...884..154W} {884, 154}

\bibitem[\protect\citeauthoryear{{Yamaguchi} et~al.,}{{Yamaguchi}
  et~al.}{2019}]{Yamaguchi2019}
{Yamaguchi} Y.,  et~al., 2019, \mn@doi [\apj] {10.3847/1538-4357/ab0d22}, \href
  {https://ui.adsabs.harvard.edu/abs/2019ApJ...878...73Y} {878, 73}

\bibitem[\protect\citeauthoryear{{Zavala} et~al.,}{{Zavala}
  et~al.}{2021}]{Zavala2021}
{Zavala} J.~A.,  et~al., 2021, \mn@doi [\apj] {10.3847/1538-4357/abdb27}, \href
  {https://ui.adsabs.harvard.edu/abs/2021ApJ...909..165Z} {909, 165}

\bibitem[\protect\citeauthoryear{{da Cunha} et~al.,}{{da Cunha}
  et~al.}{2015}]{daCunha2015}
{da Cunha} E.,  et~al., 2015, \mn@doi [\apj] {10.1088/0004-637X/806/1/110},
  \href {http://adsabs.harvard.edu/abs/2015ApJ...806..110D} {806, 110}

\makeatother
\end{thebibliography}


\appendix

\begin{table*}
\centering
\caption{HST-dark Galaxy Sample and Derived Properties}
\label{tab:sample}
\renewcommand{\arraystretch}{1.3}
\begin{tabular}{lcccccccc} 
 \hline
ID  &  RA   &  DEC   &  $z$  &  log Mass  &  log SFR  &  $A_V$  \\
    &   [deg]   &  [deg]   &    &  [$M_\odot$]  &  [$M_\odot$ yr$^{-1}$]  &  [mag]  \\[0.5ex] 
 \hline\hline
9795     &  214.9984037  &  53.0046235   &  8.06$^{+0.13}_{-0.10}$  &  2.30$^{+0.10}_{-0.09}$  &  10.24$^{+0.11}_{-0.09}$  &  2.69$^{+0.18}_{-0.17}$    \\
18038    &  214.8147216  &  52.8036165   &  8.05$^{+0.17}_{-0.11}$  &  1.99$^{+0.09}_{-0.08}$  &  9.94$^{+0.10}_{-0.09}$  &  1.77$^{+0.13}_{-0.15}$    \\
2921     &  215.0084889  &  52.9779736   &  8.01$^{+0.11}_{-0.08}$  &  1.61$^{+0.08}_{-0.07}$  &  9.54$^{+0.08}_{-0.08}$  &  1.63$^{+0.14}_{-0.12}$    \\
7194     &  214.9272387  &  52.9338962   &  7.94$^{+0.08}_{-0.06}$  &  1.90$^{+0.06}_{-0.06}$  &  9.88$^{+0.10}_{-0.08}$  &  1.64$^{+0.09}_{-0.10}$    \\
4836     &  214.9233733  &  52.9255929   &  7.28$^{+0.05}_{-0.07}$  &  1.86$^{+0.06}_{-0.07}$  &  9.79$^{+0.08}_{-0.07}$  &  2.00$^{+0.09}_{-0.10}$    \\
32479    &  214.7944062  &  52.8373983   &  7.07$^{+0.07}_{-0.04}$  &  1.68$^{+0.08}_{-0.06}$  &  9.71$^{+0.15}_{-0.13}$  &  1.46$^{+0.11}_{-0.07}$    \\
17951    &  214.8867950  &  52.8553796   &  6.83$^{+0.07}_{-1.33}$  &  2.22$^{+0.15}_{-0.13}$  &  10.46$^{+0.14}_{-0.23}$  &  2.37$^{+0.23}_{-0.18}$    \\
15222    &  214.8493871  &  52.8118241   &  6.55$^{+0.20}_{-0.33}$  &  1.60$^{+0.10}_{-0.13}$  &  10.03$^{+0.09}_{-0.09}$  &  1.49$^{+0.13}_{-0.09}$    \\
29279    &  214.8383981  &  52.8851900   &  5.99$^{+0.35}_{-0.17}$  &  1.54$^{+0.08}_{-0.07}$  &  9.47$^{+0.09}_{-0.08}$  &  1.74$^{+0.06}_{-0.07}$    \\
23985    &  214.8644947  &  52.8709797   &  5.36$^{+0.07}_{-0.05}$  &  2.06$^{+0.09}_{-0.07}$  &  10.01$^{+0.10}_{-0.09}$  &  2.13$^{+0.14}_{-0.09}$    \\
23008    &  214.8588182  &  52.8620152   &  5.18$^{+0.02}_{-0.01}$  &  1.44$^{+0.20}_{-0.16}$  &  9.36$^{+0.20}_{-0.15}$  &  3.19$^{+0.57}_{-0.42}$    \\
11011    &  214.9319470  &  52.9589227   &  5.17$^{+0.05}_{-0.05}$  &  0.91$^{+0.05}_{-0.06}$  &  8.86$^{+0.07}_{-0.07}$  &  1.14$^{+0.10}_{-0.08}$    \\
25718    &  214.8400370  &  52.8606498   &  4.72$^{+0.93}_{-0.13}$  &  1.33$^{+0.11}_{-0.19}$  &  9.60$^{+0.12}_{-0.19}$  &  1.66$^{+0.34}_{-0.25}$    \\
29538    &  214.7665813  &  52.8315228   &  4.59$^{+0.16}_{-0.19}$  &  1.10$^{+0.11}_{-0.10}$  &  9.58$^{+0.06}_{-0.09}$  &  1.74$^{+0.18}_{-0.15}$    \\
11555    &  214.9155599  &  52.9484970   &  4.47$^{+0.13}_{-0.14}$  &  0.90$^{+0.08}_{-0.10}$  &  9.59$^{+0.05}_{-0.05}$  &  0.77$^{+0.12}_{-0.13}$    \\
22858    &  214.8475516  &  52.8533710   &  4.36$^{+1.68}_{-0.52}$  &  1.60$^{+0.42}_{-0.36}$  &  10.03$^{+0.34}_{-0.35}$  &  3.95$^{+0.50}_{-0.47}$    \\
24502    &  214.8081702  &  52.8322122   &  4.22$^{+0.12}_{-0.13}$  &  1.54$^{+0.10}_{-0.10}$  &  10.10$^{+0.05}_{-0.07}$  &  2.07$^{+0.12}_{-0.10}$    \\
30673    &  214.8482898  &  52.8847859   &  3.84$^{+0.24}_{-2.09}$  &  1.06$^{+0.28}_{-0.97}$  &  9.50$^{+0.20}_{-0.45}$  &  2.01$^{+0.61}_{-0.28}$    \\
9594     &  214.9508404  &  52.9668643   &  3.74$^{+0.46}_{-1.99}$  &  0.91$^{+0.21}_{-0.95}$  &  9.51$^{+0.17}_{-0.70}$  &  2.28$^{+0.79}_{-0.28}$    \\
9784     &  214.9091114  &  52.9372131   &  3.65$^{+1.95}_{-1.90}$  &  1.39$^{+0.64}_{-0.96}$  &  9.95$^{+0.46}_{-0.63}$  &  3.26$^{+1.69}_{-0.41}$    \\
27303    &  214.7680265  &  52.8164017   &  3.56$^{+0.13}_{-0.15}$  &  2.14$^{+0.14}_{-0.14}$  &  10.68$^{+0.09}_{-0.12}$  &  3.34$^{+0.18}_{-0.20}$    \\
16342    &  214.8871236  &  52.8453781   &  3.56$^{+0.69}_{-2.02}$  &  0.72$^{+0.27}_{-0.97}$  &  9.30$^{+0.29}_{-0.68}$  &  2.06$^{+0.73}_{-0.39}$    \\
20994    &  214.8556793  &  52.8487133   &  3.50$^{+0.20}_{-1.74}$  &  0.33$^{+0.17}_{-0.65}$  &  8.96$^{+0.09}_{-0.38}$  &  1.48$^{+0.43}_{-0.19}$    \\
24835    &  214.8894314  &  52.8922071   &  3.30$^{+0.10}_{-0.15}$  &  0.43$^{+0.10}_{-0.15}$  &  8.92$^{+0.10}_{-0.11}$  &  0.87$^{+0.17}_{-0.14}$    \\
2439     &  214.9440430  &  52.9297453   &  2.04$^{+1.79}_{-0.95}$  &  0.38$^{+0.73}_{-0.83}$  &  9.20$^{+0.32}_{-0.53}$  &  2.68$^{+1.10}_{-0.44}$    \\
22807    &  214.8476086  &  52.8534052   &  1.86$^{+2.01}_{-0.23}$  &  0.32$^{+0.95}_{-0.20}$  &  9.15$^{+0.62}_{-0.23}$  &  4.53$^{+0.61}_{-1.18}$    \\
21771    &  214.8560277  &  52.8546738   &  1.76$^{+1.85}_{-0.11}$  &  0.81$^{+1.05}_{-0.22}$  &  9.67$^{+0.35}_{-0.17}$  &  3.63$^{+0.45}_{-0.61}$    \\
4710     &  215.0045567  &  52.9835262   &  1.73$^{+1.86}_{-0.20}$  &  1.02$^{+0.98}_{-0.25}$  &  9.99$^{+0.62}_{-0.19}$  &  4.40$^{+0.49}_{-1.16}$    \\
28170    &  214.7598238  &  52.8334124   &  1.68$^{+1.90}_{-0.12}$  &  -0.09$^{+0.88}_{-0.22}$  &  8.80$^{+0.50}_{-0.22}$  &  3.33$^{+0.45}_{-0.69}$    \\
10881    &  214.9257706  &  52.9544456   &  1.32$^{+0.72}_{-0.22}$  &  0.14$^{+0.54}_{-0.31}$  &  9.14$^{+0.40}_{-0.23}$  &  4.04$^{+0.77}_{-0.80}$    \\
\hline 
\end{tabular}

\end{table*}




\bsp	
\label{lastpage}
\end{document}